\documentclass[twocolumn,traditabstract]{aa}  
\usepackage{graphicx}
\usepackage{txfonts}
\usepackage{amsmath,amsfonts,amssymb}
\usepackage{fixltx2e}
\usepackage{caption,subcaption}
\usepackage[breaklinks, colorlinks, citecolor=blue]{hyperref}
\usepackage{color}
\usepackage{natbib}

\usepackage{tabularx}

\bibpunct{(}{)}{;}{a}{}{,}

\def\setsymbol#1#2{\expandafter\def\csname #1\endcsname{#2}}
\def\getsymbol#1{\csname #1\endcsname}

\newbox\tablebox    \newdimen\tablewidth
\def\leaderfil{\leaders\hbox to 5pt{\hss.\hss}\hfil}
\def\endPlancktable{\tablewidth=\columnwidth 
    $$\hss\copy\tablebox\hss$$
    \vskip-\lastskip\vskip -2pt}

\def\tablenote#1 #2\par{\begingroup \parindent=0.8em
    \abovedisplayshortskip=0pt\belowdisplayshortskip=0pt
    \noindent
    $$\hss\vbox{\hsize\tablewidth \hangindent=\parindent \hangafter=1 \noindent
    \hbox to \parindent{$^#1$\hss}\strut#2\strut\par}\hss$$
    \endgroup}
\def\doubleline{\vskip 3pt\hrule \vskip 1.5pt \hrule \vskip 5pt}

\newcommand{\Planck}{{\em Planck}}
\newcommand{\Euclid}{{\em Euclid}}
\newcommand{\XMM}{XMM-{\em Newton}}
\newcommand{\PSZ}{PSZ1}

\newcommand{\Mfive}{M_{500}}

\newcommand{\Msz}{M_{\rm SZ}}

\newcommand{\Mcl}{M_{\rm CL}}
\newcommand{\Ml}{M_{\rm L}}
\newcommand{\Mpl}{M_{\rm PL}}
\newcommand{\zspec}{z_{\rm spec}}
\newcommand{\Nc}{N_{\rm clus}}

\newcommand{\sigl}{\sigma_{\rm L}} 

\newcommand{\sigsz}{\sigma_{\rm SZ}} 

\newcommand{\Mcut}{M_{\rm cut}}

\newcommand{\sigerf}{\sigma_{\rm cut}}

\newcommand{\asz}{\alpha_{\rm SZ}}
\newcommand{\bsz}{b_{\rm SZ}}
\newcommand{\al}{\alpha_{\rm L}}
\newcommand{\bl}{b_{\rm L}}

\newcommand{\bp}{\vec{p}}

\newcommand{\zmin}{{z_{\rm min}}}
\newcommand{\zmax}{{z_{\rm max}}}

\newcommand{\LCDM}{\hbox{$\Lambda$CDM}}

\newcommand{\msun}{\mbox{M}_\odot}

\newcommand{\avg}[1]{\langle #1 \rangle}

\newcommand{\be}{\begin{equation}}
\newcommand{\ee}{\end{equation}}

\newcommand{\bea}{\begin{eqnarray}}
\newcommand{\eea}{\end{eqnarray}}

\newcommand{\nn}{\nonumber}

\newcommand{\lkhd}{{\cal{L}}}

\newcommand{\bM}{\vec{M}}
\newcommand{\bz}{\vec{z}}

\newcommand{\ndet}{n_{\rm det}}

\begin{document}

   \title{Calibrating the \Planck\ cluster mass scale with CLASH}


\author{M. Penna-Lima
\inst{\ref{inst1},\ref{inst2}}
\and
J.~G. Bartlett
\inst{\ref{inst1},\ref{inst3}}
\and
E. Rozo
\inst{\ref{inst4}}
\and
J.-B. Melin
\inst{\ref{inst5}}
\and
J. Merten
\inst{\ref{inst6}}
\and
A.~E. Evrard
\inst{\ref{inst7},\ref{inst8}}
\and
M. Postman
\inst{\ref{inst9}}
\and
E. Rykoff
\inst{\ref{inst10},\ref{inst11}}
}

\institute{
APC, AstroParticule et Cosmologie, Universit\'e Paris Diderot, CNRS/IN2P3, CEA/Irfu, Observatoire de Paris, Sorbonne Paris Cit\'e, 10, rue Alice Domon et L\'eonie Duquet, 75205 Paris Cedex 13, France\\
\email{pennal@apc.in2p3.fr}\label{inst1}
\and
Centro Brasileiro de Pesquisas Físicas, Rua Dr. Xavier Sigaud 150, Rio de Janeiro, 22290-180, RJ, Brazil
\label{inst2}
\and
Jet Propulsion Laboratory, California Institute of Technology, Pasadena, California, USA
\label{inst3}
\and
Department of Physics,  University of Arizona,  Tucson,  AZ
85721, USA
\label{inst4}
\and
DRF/Irfu/SPP, CEA-Saclay, F-91191 Gif-sur-Yvette Cedex, France
\label{inst5}
\and
Department of Physics, University of Oxford, Keble Road, Oxford OX1 3RH, UK
\label{inst6}
\and
Deparment of Physics and Michigan Center for Theoretical Physics, University of Michigan, Ann Arbor, MI 48109 USA
\label{inst7}
\and
Department of Astronomy, University of Michigan, Ann, Arbor, MI 48109 USA
\label{inst8}
\and
Space Telescope Science Institute, 3700 San Martin Drive, Baltimore, MD 21208, USA
\label{inst9}
\and
Kavli Institute for Particle Astrophysics \& Cosmology, P.O. Box 2450, Stanford University, Stanford, CA 94305, USA
\label{inst10}
\and
SLAC National Accelerator Laboratory, Menlo Park, CA 94025, USA
\label{inst11}
}

   \date{Received 28 October 2016 / Accepted 12 June 2017}
 
  \abstract{We determine the mass scale of \Planck\ galaxy clusters using gravitational lensing mass measurements from the Cluster Lensing And Supernova survey with Hubble (CLASH).  We have compared the lensing masses to the \Planck\ Sunyaev-Zeldovich (SZ) mass proxy for 21 clusters in common, employing a Bayesian analysis to simultaneously fit an idealized CLASH selection function and the distribution between the measured observables and true cluster mass. We used a tiered analysis strategy to explicitly demonstrate the importance of priors on weak lensing mass accuracy.  In the case of an assumed constant bias, $\bsz$, between true cluster mass, $\Mfive$, and the \Planck\ mass proxy, $\Mpl$, our analysis constrains $1-\bsz = 0.73\pm 0.10$ when moderate priors on weak lensing accuracy are used, including a zero-mean Gaussian with standard deviation of 8\% to account for possible bias in lensing mass estimations. Our analysis explicitly accounts for possible selection bias effects in this calibration sourced by the CLASH selection function. Our constraint on the cluster mass scale is consistent with recent results from the Weighing the Giants program and the Canadian Cluster Comparison Project. It is also consistent, at 1.34$\sigma$, with the value needed to reconcile the \Planck\ SZ cluster counts with \Planck's base $\Lambda$CDM model fit to the primary cosmic microwave background anisotropies.
}
\keywords{Galaxies: clusters: general; Cosmology: observations ; cosmological parameters; dark matter
               }

   \maketitle
%

\section{Introduction}
Galaxy cluster mass measurements are the dominant source of systematic uncertainty in cosmological constraints derived from the space-time abundance of galaxy clusters. This was acutely illustrated by the \Planck\ collaboration's finding of tension between the \LCDM\ cosmology parameters favored by cluster counts and those derived by combining the primary cosmic microwave background (CMB) anisotropies with non-cluster data \citep[PXX][]{PlanckXX1303.5080,PlanckXXIV1502.01597}.  While the discrepant findings could reflect a relatively large neutrino mass or more exotic physics, the confidence in such statements is limited by systematic uncertainties in mass measurements \citep{Rozo1302.5086}.  

The fundamental issue is that cluster halo mass is not directly observable.  While N-body simulations have calibrated the space density of massive halos to good precision \citep[e.g.,][and references therein]{Bhattacharya1005.2239, Murray1306.6721}, application to cluster counts on the sky requires the use of scaling relations between halo mass and observable cluster properties, often termed mass proxies. In the case of \Planck\ clusters detected through the Sunyaev-Zeldovich (SZ) effect, the required relation is between SZ signal strength and mass.  

In principle, scaling relations can be calibrated with hydrodynamic simulations that properly account for most relevant physical processes.  The approach is currently limited, however, by uncertainties in the baryon physics associated with galaxy feedback mechanisms \citep{Ragone-Figueroa2013, Dubois2013, LeBrun2014, Martizzi2014, Genel2014}.  Nevertheless, the fidelity of the simulations is high enough to provide insights into the general form of the scaling relations and into important sources of systematic error.

In practice, empirical approaches are used to establish scaling relations.  The \Planck\ analysis employed X-ray observations from \XMM\ to derive masses based on the assumption of hydrostatic equilibrium (HSE) of the intra-cluster medium (ICM) \citep{Arnaud0910.1234}.  For decades, hydrodynamic simulations have indicated that HSE is not exact, with expectations that HSE masses underestimate true values by typically tens of percent, depending on scale and the exact modeling of baryon physics \citep[e.g.,][]{Evrard1990, Rasia2006, Nagai2007, Piffaretti2008, Rasia2012, Battaglia2013, Nelson1308.6589}. Combining \XMM\ and {\it Chandra} data, \citet{Mahdavi2013} found a 15\%  systematic difference in HSE masses. Moreover, the X-ray instrument calibration errors can also affect masses at the ten-percent level \citep{Donahue1405.7876,Rozo2014a}.  

Gravitational lensing offers a powerful, independent alternative to HSE masses, but individual weak lensing mass estimates are noisy due to the sensitivity of the broad lensing kernel to material along the line-of-sight \citep{Hoekstra2003} and to halo triaxiality \citep{Becker2011}.  Detailed lensing image simulations by \citet{Meneghetti2010a} found that weak lensing mass estimates incur smaller bias than X-ray HSE values, with mean biases of a few percent.   Independent work by \citet{Becker2011} supports this result, but mean underestimates of up to 10\% have also been reported \citep{Rasia2012}.  The bias depends in part on the method of extracting cluster mass from the lensing data, so precise estimates of systematic error require careful modeling of the full data acquisition and analysis workflow.
 
The \Planck\ cluster cosmology findings motivate deeper investigation into the mass calibration for that sample.  Here, we have relied on recent results from the Cluster Lensing and Supernova survey with Hubble \citep[CLASH, ][]{Postman2012} collaboration, who have measured lensing masses from imaging data of exquisite depth and wavelength coverage for a sample of 25 X-ray and lensing selected galaxy clusters \citep{Merten2015,Umetsu2014, Zitrin2015}. In particular, we used the  reconstruction method presented in \citep{Merten2015} to obtain 22 CLASH mass estimates.\footnote{There is not enough data for the remaining three CLASH clusters, namely, MACS0647+7015, MACS2129-0741 and Abell1423. For instance, there is no wide-field 3+-band Subaru data available for the cluster Abell1423.} After cross-matching the CLASH and \Planck\ cluster catalogs, we measured the SZ signal of 22 CLASH clusters in the publicly available \Planck\ dataset for any CLASH clusters not included in the original \Planck\ cluster catalog.   We then used the CLASH cluster sample to place tight constraints on the SZ scaling relation of galaxy clusters, and discuss the import of our results for the cosmological interpretation of the \Planck\ cluster counts.  

The layout of the paper is as follows.  In Sect.~\ref{sec:catalogs} we summarize the methods 
used to compute the CLASH and \Planck\ mass estimates for the 21 clusters in common, while in Sect.~\ref{sec:mass_comparison} we compare these estimates. In Sect.~\ref{sec:analysis} we introduce our models for the CLASH selection function and for the measurements, that is, the mass-observable distribution, including both observational uncertainties and intrinsic covariance between lensing and SZ signals.  We construct the posterior probability distribution and perform different Bayesian analyses in Sect.~\ref{sec:results} to constrain the model parameters. In Sects.~\ref{sec:discussion} and \ref{sec:conclusion} we discuss our results and then conclude with final remarks.  

Unless otherwise specified, we have adopted a fiducial flat cosmology with $\Omega_{\rm M} = 0.3 = 1 - \Omega_\Lambda$ (see Table~\ref{tab:cosmo_params}), and all masses are given within $R_{500}$, the radius at which the mean mass density within the cluster reaches 500 times the critical density at the redshift of the cluster: $M = (4\pi/3)R^3_{500} (500 \rho_{\rm c})$, with $\rho_{\rm c} = 3H^2(z)/8\pi G$.  We refer to the \Planck\ mass proxy (see below) as $\Mpl$ and to the CLASH lensing mass as $\Mcl$.  These measurements are noisy realizations of the true SZ mass proxy, $\Msz$, and true lensing mass, $\Ml$.


\section{Masses and mass proxies}
\label{sec:catalogs}
\subsection{CLASH lensing masses}
The CLASH survey \citep{Postman2012} is a Hubble Space Telescope (HST) multi-cycle treasury program targeting 25 massive galaxy clusters in the redshift range $0.18 < z < 0.89$ and over a mass range $0.5\times10^{15}\ h^{-1} \msun \lesssim M_{\textrm{vir}} \lesssim 2.0\times10^{15}\ h^{-1} \msun$.  The sample of 25 clusters was further subdivided into an X-ray selected sub-sample (20 clusters), and a strong-lensing selected sub-sample (five clusters, also referred to as the high-magnification sub-sample). For a complete definition of the two sub-sets, see \citet{Postman2012}.

Each cluster was observed for 20 HST orbits in 16 broad photometric passbands.  These data are supplemented with a three-to-five band Subaru/Suprime-Cam and ESO/WFI optical imaging to enable weak lensing measurements of the cluster profiles out to their virial radii. The combination of HST and ground-based wide-field data allows for a comprehensive weak-and strong-lensing analysis.

We have used a total of 22 CLASH clusters, of which 19 belong to the X-ray selected sub-sample and three 
belong to the high-magnification subset. The lensing reconstructions of the X-ray selected clusters 
have recently been presented in \citet{Merten2015} and \citet{Umetsu2014}, while mass estimates for 
the high-magnification clusters have in part been presented in \citet{Medezinski2013} and 
\citet{Umetsu2014}. Recently, \citet{Umetsu2016} and \citet{Zitrin2015} reconstructed the surface 
mass density profiles of 20 and 25 
clusters (complete CLASH sample), respectively. In all cases, a combination of weak and strong lensing 
was used to derive reliable masses. 

A thorough description of the input data and reconstruction techniques used in this work, is given by \citet{Merten2015}.  Here, we only provide a brief summary.
Masses were derived with the \texttt{SaWLens} code \citep{Merten2009} that consistently combines weak and strong lensing
in a non-parametric fashion on adaptively refined grids. The method was thoroughly tested with realistic lensing scenarios
and numerically simulated clusters \citep{Meneghetti2010a, Rasia2012}, and has been used multiple times for the reconstruction 
of real galaxy clusters \citep{Merten2009, Merten2011, Umetsu2012, Patel2014}. For the CLASH analysis, \texttt{SaWLens} combines constraints from HST strong lensing, HST weak lensing and wide-field ground-based weak lensing (except for CLJ1226+3332 whose mass was reconstructed using HST data only) into a single reconstruction of the cluster's gravitational potential from which it derives the surface-mass density. NFW fits to the surface-mass density provide the desired total 3-dimensional mass of the halo at any given radius. 

Error bars for the lensing reconstruction were derived from 1000 bootstrap resamplings of the input weak-lensing shear catalogs,  including their photometric redshift uncertainties,\footnote{During the bootstrap the redshift uncertainty of a given WL galaxy was also sampled. Then the angular distance ratio $D_s/D_{ls}$ for each individual galaxy was calculated, which was then averaged and used to convert the convergence map into a physical surface-density map.
} and by randomly sampling the allowed redshift range of strong-lensing multiple image systems. Candidate systems are treated by random
inclusion or exclusion in each bootstrap. From the bootstrap realizations, we have derived the covariance matrix of the surface-mass density
bins, which is then taken into account during the NFW profile fitting \citep[see][for details]{Merten2015}. 

In particular, as discussed in \citet{Merten2015}, the mass estimate of the system CLJ1226+3332 shows only a mild tension within the errors with an independent study by \citet{Jee2009}. The uncertainties in the source redshift distribution are taken into account by the error estimation of the SaWLens method, which is based on a combination of bootstrapping and a resampling of source redshifts within their uncertainties.

These CLASH mass errors account for a number of systematics in the lensing reconstruction, including shape scatter in the weak lensing catalogs,
redshift uncertainties in the weak lensing background and strong-lensing multiple-image populations, mis-identifications of strong-lensing
features and uncertainty in the central peak position of the mass distribution. Sources of systematic error not covered
by this analysis include correlated and un-correlated structure within the cluster field, effects of tri-axiality and general error stemming from the fact that we fit a simplified 1D analytical form of the density profile to a complex mass distribution. These unaccounted sources of systematic error in CLASH mass 
estimates have recently been estimated by \citet{Umetsu2014} to be $\sim$8\%, and a detailed comparison 
of the SaWLens mass reconstructions to a set of numerical simulations that mimic the CLASH selection 
functions has been presented in \citet{Meneghetti2014}. In Sect.~\ref{sec:results} we will account for this systematic uncertainty using a Gaussian distribution with standard deviation of 8\% \citet{Umetsu2014}, and also the left-skewed distribution for the 3D and true mass ratio found by  \citet{Meneghetti2014}, as priors on the lensing mass scale.


\begin{figure*}
\centering
\includegraphics[width=.45\textwidth]{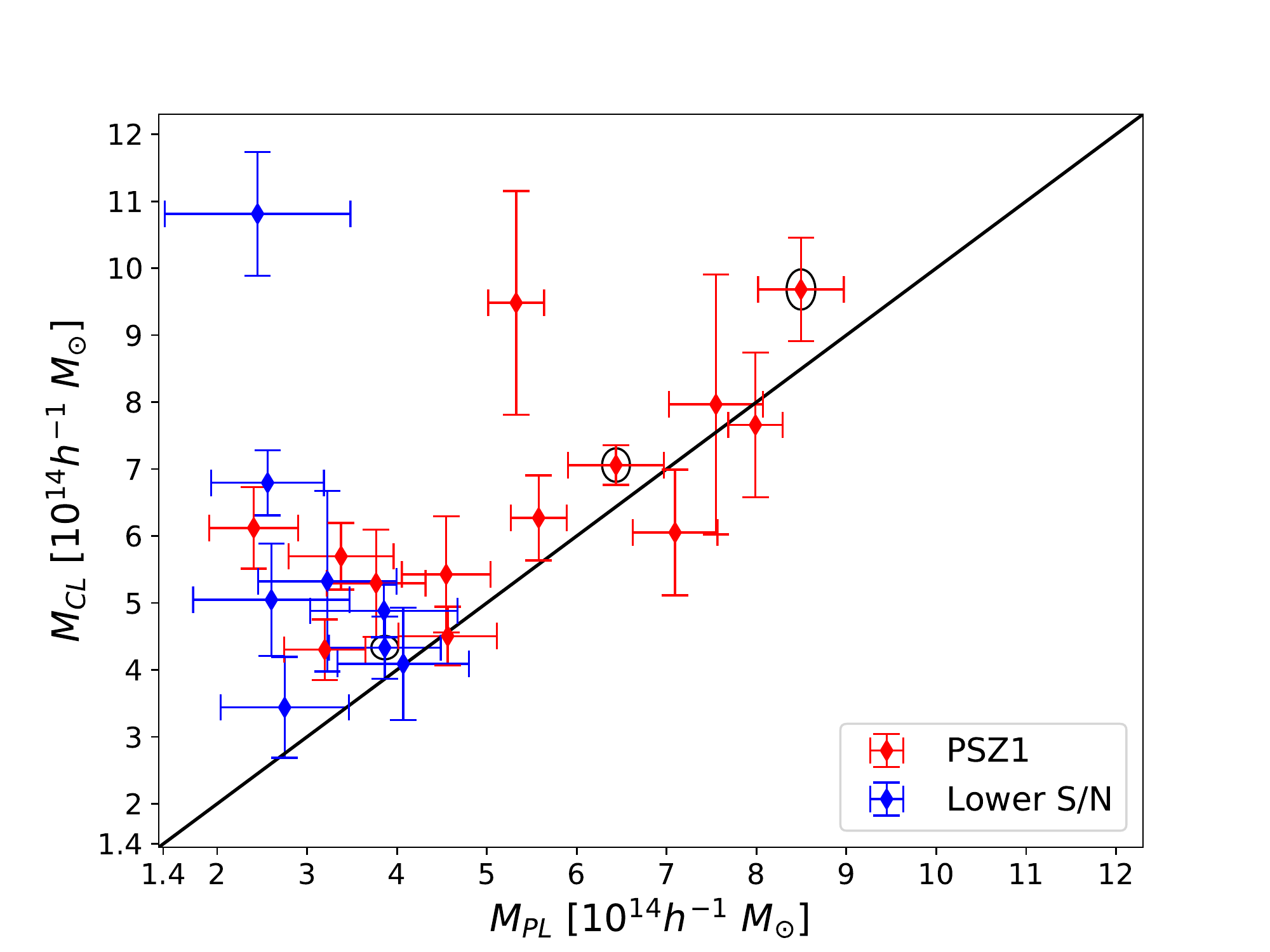}
\includegraphics[width=.45\textwidth]{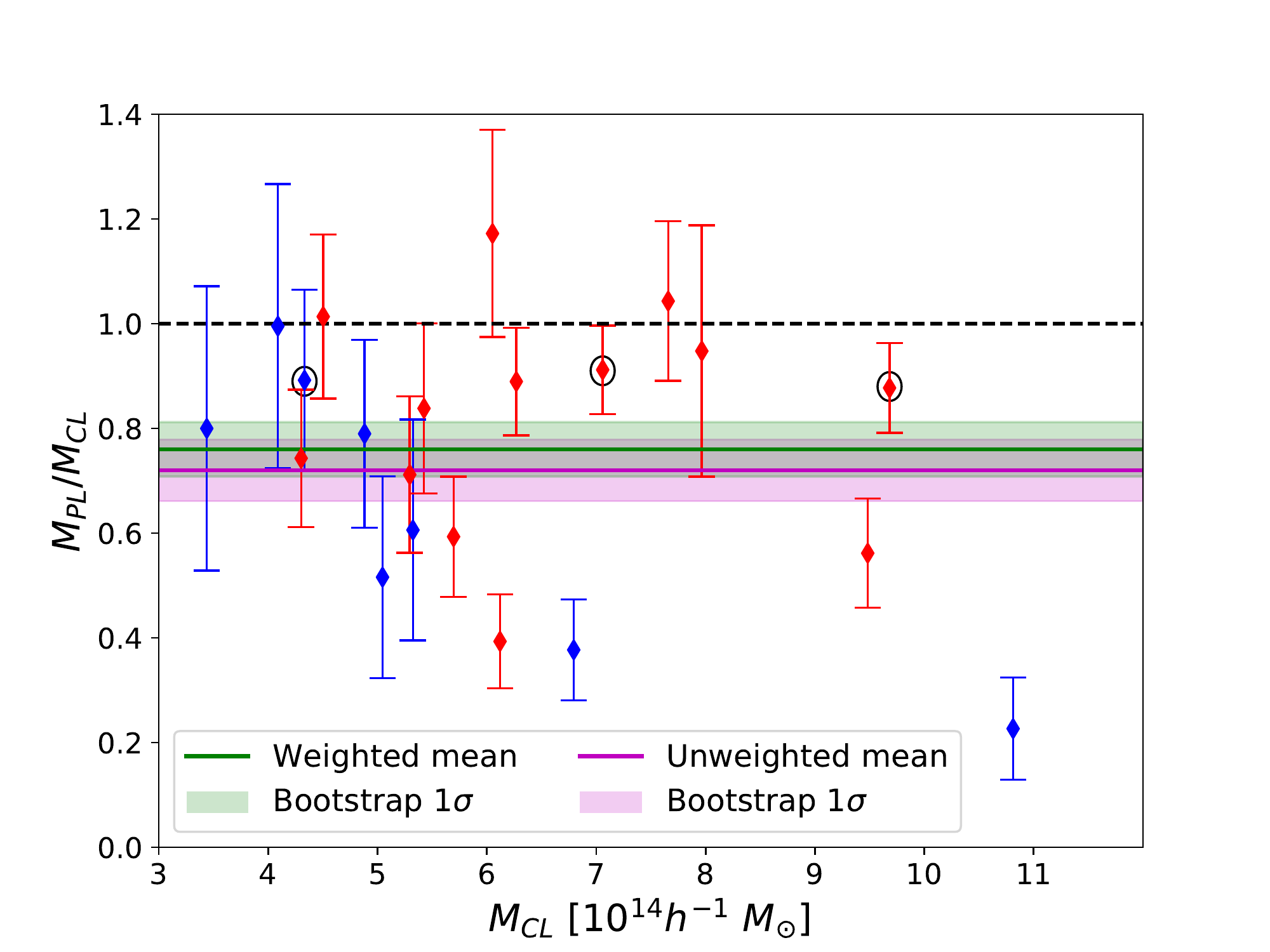}
\caption{Mass comparison and mass bias estimates.  On the {\em left}, we plot the CLASH lensing masses as a function of the \Planck\ SZ mass proxy.  Red points represent clusters found in the PSZ1 (13 objects) and blue those of the remaining objects at lower signal-to-noise (eight clusters); open circles identify the three clusters in the high-magnification subsample.  The solid line is the equality line. The strongest outlier in the upper left corner is CLJ1226 at $z=0.89$, the highest redshift object in the sample.  The {\em right-hand panel} plots $\rho = \Mpl / \Mcl$ as a function of CLASH mass together with its uncertainty $\Delta\rho$ (see text) in the same color scheme.  
The green line and band correspond to the sample mean, $\avg{\rho}_{\rm s}$, and its standard deviation obtained from a bootstrap analysis: $E(\avg{\rho}_{\rm s}) = 0.72 \pm 0.059$, where $E$ indicates a calculation over the bootstrap and  $\avg{\rho}_{\rm s}$ is calculated as the inverse-variance weighted mean.  Similarly, the magenta line and band represent the bootstrap mean and uncertainty of the unweighted mean: $E(\avg{\rho}_{\rm s}) = 0.76 \pm 0.052$. The dashed line indicates zero mass bias. \\}
\label{fig:data}  
\end{figure*} 



\subsection{\Planck\ SZ mass proxy}
\label{sec:Mpl}
We attempted to measure the SZ mass proxy for each of the 22 CLASH clusters.  Of these, 13 are found in the \Planck\ Catalog of SZ Sources \citep[PSZ1,][]{psz1} and an additional eight are detected in the \Planck\ temperature maps, albeit at lower significance than for clusters in
the \PSZ.  The remaining system, MACSJ1311-03, lies in a dusty region of the sky and has a negative signal-to-noise ratio in the \Planck\ data.  It is unusable, and we remove this system from our analysis, leaving 21 clusters.

The SZ mass  proxy for each cluster is extracted following the same procedure as for the \PSZ.  Here, we only provide a brief summary of the method. 
For each CLASH cluster, we extract $10 \deg \times 10 \deg$ tangential maps centered on the cluster position for each of the six \Planck\ High Frequency Instrument (HFI) channels.  The maps were filtered with the SZ Matched MultiFilter \citep[MMF3,][]{melin2006}, varying the characteristic scale, $\theta_s$, of the filter.  For each system, the filter provided a degeneracy curve relating the SZ signal strength, $Y_{500}$, to the characteristic scale, $\theta_s$. We break this degeneracy curve with an X-ray prior on the signal-size relation \citep[Eq. 20 of][]{PlanckXX1303.5080}, obtaining the cluster signal strength, $Y_{500}$, and scale, $\theta_s$, independently, and then convert the latter to $\Mpl$. The upper and lower values for $\Mpl$ are obtained following the same method, but using $Y_{500} \pm \sigma_{Y_{500}}$ versus $\theta_s$ for the degeneracy curve, where $\sigma_{Y_{500}}$ is the error on the signal provided by the MMF at each characteristic scale.  Details are given in Sect. 7.2.2 of~\cite{psz1}.
 
We note that, in this work, the \Planck\ masses were derived when centering the filters on the fiducial CLASH cluster position, irrespective of the \PSZ\ location.  This reflects the fact that \Planck\ positions are more uncertain than those from the CLASH catalog.  Because of this, our new SZ measurements differ slightly from, but remain consistent with, the values published in the \PSZ.  
The inverse-variance weighted average of the ratio of the masses inferred from the SZ between the \PSZ\ 
and our CLASH-centered estimates is $0.99 \pm 0.03$.

\begin{table}
\caption{Cluster redshifts, \Planck\ and CLASH mass estimates.}             
\label{table:catalog}      
\centering             
\renewcommand{\arraystretch}{1.5}   
\resizebox{\columnwidth}{!}{
\begin{tabular}{c | c | c | c  }  
\hline\hline                 
{Cluster}& $z$ & $\Mpl [10^{14} h^{-1} \msun]$ & $\Mcl [10^{14} h^{-1} \msun]$  \\
\hline                       
Abell-383 & 0.188 & $2.41 \pm 0.49$ & $6.12 \pm 0.61$ \\ 
Abell-2261 & 0.225 & $5.33 \pm 0.31$ & $9.48 \pm 1.67$ \\ 
MACSJ1206-08 & 0.439 & $7.09 \pm 0.47$ & $6.05 \pm 0.94$ \\
RXJ1347-1145 & 0.451 & $7.55 \pm 0.52$ & $7.97 \pm 1.94$ \\ 
MACSJ0329-02 & 0.45 &  $2.60 \pm 0.87$ & $5.05 \pm 0.84$ \\ 
MS2137-2353 & 0.313 & $2.56 \pm 0.63$ & $6.79 \pm 0.49$ \\ 
MACSJ0744+39 & 0.686 & $3.86 \pm 0.82$ & $4.88 \pm 0.39$ \\ 
MACSJ1115+0129 & 0.352 & $4.55 \pm 0.49$ & $5.43 \pm 0.87$ \\
Abell-611 & 0.288 & $3.38 \pm 0.58$ & $5.70 \pm 0.50$ \\
RXJ1532.8+3021 & 0.363 & $2.75 \pm 0.71$ & $3.44 \pm 0.75$ \\
MACSJ1720+3536 & 0.391 & $3.77 \pm 0.55$ & $5.29 \pm 0.80$ \\
RXJ2129+0005 & 0.234 & $3.20 \pm 0.45$ & $4.30 \pm 0.45$ \\
MACSJ1931-26 & 0.352 & $4.56 \pm 0.55$ & $4.50 \pm 0.44$ \\
Abell-209 & 0.206 & $5.58 \pm 0.31$ & $6.27 \pm 0.64$ \\
RXCJ2248-4431 & 0.348 & $7.99 \pm 0.30$ & $7.66 \pm 1.08$ \\
MACSJ0429-02 & 0.399 & $3.23 \pm 0.77$ & $5.32 \pm 1.35$ \\
MACSJ1423+24 & 0.545 & $4.07 \pm 0.73$ & $4.09 \pm 0.84$ \\
CLJ1226+3332 & 0.89 & $2.45 \pm 1.03$ & $10.81 \pm 0.92$ \\
MACSJ0717+37 & 0.548 & $8.49 \pm 0.48$ & $9.68 \pm 0.77$ \\
MACSJ1149+22 & 0.544 & $6.44 \pm 0.53$ & $7.06 \pm 0.30$ \\
MACSJ0416-24 & 0.42 &  $3.87 \pm 0.62$ & $4.33 \pm 0.46$ \\
\hline                  
\end{tabular}}
\end{table}


\section{Mass comparison}
\label{sec:mass_comparison}
The left-hand panel of Fig.~\ref{fig:data} compares the CLASH masses to the \Planck\ SZ mass proxy, with the red points representing clusters in the \PSZ\ and blue the additional lower signal-to-noise clusters.  The three circled data points identify the CLASH clusters selected for their high lensing magnification. The redshift value and both \Planck\ and CLASH mass estimates of each cluster are summarized in Table~\ref{table:catalog}.  

The CLASH lensing masses tend to be larger than the \Planck\ proxy values, and there is a ``wall'' of clusters at low \Planck\ mass ($\Mpl=2\times 10^{14}\ h^{-1}\msun$) reaching to high lensing masses and that would appear to be related to the CLASH selection function. Our subsequent analysis indeed finds an effective mass cutoff for CLASH selection around $(4-5) \times 10^{14}\ h^{-1}\msun$.
The biggest outlier, in the upper left corner in blue, is CLJ1226 at \mbox{$z=0.89$}, the highest redshift object in the CLASH sample.

Following other work \citep[e.g.,][]{vonderLinden2014, Hoekstra2015, Simet2017}, we began by assuming that the sample mean mass ratio, $\avg{\Mpl/\Mcl}_s$, is an unbiased estimator of the mass bias, $(1-\bsz)$, between the \Planck\ mass and true halo mass, $\Mfive$.  The right-hand panel of Fig.~\ref{fig:data} shows the mass ratio $\rho = \Mpl/\Mcl$ for each of the clusters. Its uncertainty is calculated as $(\Delta (\ln\rho))^2 = (\Delta (\ln\Mpl))^2 + (\Delta (\ln\Mcl))^2$ and no correlation is assumed for the \Planck\ and CLASH mass errors.

We first calculated the inverse-variance weighted mean: $\avg{\rho}_s = 0.72\pm 0.057$.  We also estimate the uncertainty with a bootstrap, using the \textsf{boot} function from the Bootstrap R package \citep{boot}, to obtain $E(\avg{\rho}_s) = 0.72 \pm 0.059$, where $E$ indicates a calculation over the bootstrap.  The uncertainties on the cluster mass ratios are at best approximate, having been calculated using the measured ratios for each cluster, which tends to overweight (underweight) low (high) valued excursions and pull the inverse-variance weighted mean to smaller values. 
An unweighted mean with bootstrap errors, as per \citet{vonderLinden2014}, results in a slightly higher value of $E(\avg{\rho}_s) = 0.76 \pm 0.052$ (magenta line and band). 
These initial results are all in agreement with those obtained by the Canadian Cluster Comparison Project \citep[CCCP,][]{Hoekstra2015} and the Weighing the Giants program \citep[WtG,][]{vonderLinden2014}.

We note that three of our CLASH clusters were selected for their previously known high magnification, rather than based on their X-ray properties, as is true for the bulk of the sample; these objects are indicated in Fig.~\ref{fig:data}.  The X-ray selection would more naturally lend itself to our modeling of the selection function.  To check for any sensitivity to these objects, we also performed our bootstrap analysis without them (i.e., on the other 18, X-ray selected, clusters).  The results do not change significantly.  Curiously, when eliminating these high magnification clusters, presumably more likely to have large masses, we find a slight increase in the mass bias, namely, $E(\avg{\rho}_s) = 0.68 \pm 0.066$.  This can be appreciated by eye from Fig.~\ref{fig:data}, where these three clusters all lie close to the equality line.
 
The simple analysis performed above is not completely satisfactory for a variety of reasons.  Moreover, it is important to note that the average for this estimator is strictly over an ensemble including both measurement errors and intrinsic, potentially correlated astrophysical scatter at fixed true mass.  
\cite{Meneghetti2014} found that the intrinsic scatter in CLASH lensing-deduced masses is expected to be log-normal with a standard deviation of (10-15)\%, although it is potentially larger due to the impact of correlated structure\footnote{Recently, \citet{Shirasaki2016} reported that the scatter of about $20\%$ in the thermal SZ and weak lensing relation, found from observations \citep{Marrone2012}, is due to projections of correlated structures and the bias in the lensing determined cluster radius.}, that was not fully accounted for in the \cite{Meneghetti2014} simulations \citep[see also][]{Becker2011}.  Intrinsic scatter in the \Planck\ mass proxy is related to the scatter in SZ signal at fixed halo mass, estimated at $\sim 10\%$ according to numerical simulations \citep[e.g.,][]{Nagai2007, Stanek2010}. The exact way this propagates to the \Planck\ mass is not quantified.
One would also expect a positive correlation between the lensing and SZ signals because both are a linear projection along the line-of-sight \citep{Noh01052011, Angulo2012}.

Finally, the simple analysis above has no means of 
accounting for selection criteria in the cluster sample (especially for a rather small and peculiar
sample like CLASH), which is critical for interpreting the relation between the observed mass ratio and the mass bias of the SZ masses relative to true halo mass.  
The Bayesian approach presented in the next section aims to address these shortcomings and thereby provide a more robust result and error analysis.


\section{Bayesian analysis}
\label{sec:analysis}

The goal of our Bayesian analysis is to constrain models for the distribution of lensing and \Planck\ masses given true halo mass and, simultaneously, an approximate form of the CLASH selection function.  Our primary objective is the mass bias parameter, $(1-\bsz)$, quantifying the bias between the \Planck\ mass proxy and true halo mass.  With the Bayesian analysis, we can incorporate important astrophysical effects, such as the correlation between SZ and lensing signals, and evaluate their importance.  The first task is to construct the posterior probability distribution for the model parameters given the data.

Our data consist of a set of \Planck-\ and CLASH-determined masses and spectroscopic redshifts that we arrange into three data vectors, $\vec{\Mpl}$, $\vec{\Mcl}$ and $\vec{\zspec}$, respectively. Each vector has as many elements as clusters in our sample, $\Nc=21$.  From these data and a model for the distribution of their uncertainties, we wish to determine the true cluster masses, $\vec{\Mfive}$, and the mass bias, $(1-\bsz)$, as defined below.

\subsection{Model}
The observed \Planck\ and CLASH masses are noisy Gaussian realizations of the underlying SZ and lensing masses, denoted $\Msz$ and $\Ml$, respectively. That is, $P(\Mpl|\Msz)$ is a Gaussian distribution of mean $\Msz$ and variance given by the {\Planck} measurement error.  The same holds for $P(\Mcl|\Ml)$.  The masses $\Msz$ and $\Ml$ are in turn related to the true halo mass $\Mfive$ via a bivariate log-normal distribution.  
We took the mean of these quantities to be
\bea
\avg{\ln\Msz | \Mfive} & = & \ln(1- \bsz) + \asz \ln\left(\frac{\Mfive}{M_0}\right) , \label{eq:avgmsz} \\
\avg{\ln\Ml | \Mfive} & = & \ln(1- \bl) + \al \ln\left(\frac{\Mfive}{M_0}\right). \label{eq:avgml}
\eea
In these expressions, it is understood that $\Msz$ and $\Ml$ are in units of the pivot mass, $M_0$.  The intrinsic scatter in $\ln\Msz$ ($\ln\Ml$) at fixed mass is denoted via $\sigsz$ ($\sigl$), the correlation coefficient between the two scatters is $r_{\rm{SZ, L}\vert \Mfive}$ (but for simplicity we denote it by $r$), and we adopt a pivot mass as the median of the CLASH lensing masses, i.e., $M_0 = 5.7 \times 10^{14} \, h^{-1} \msun$. 

The parameter $(1-\bsz)$ in Eq.~(\ref{eq:avgmsz}) is the mass bias we seek to calibrate the mass scale of the \Planck\ clusters.  It accounts for any source of bias, instrumental (e.g., X-ray satellite calibration) or astrophysical (e.g., violation of hydrostatic equilibrium in the ICM).  Although defined here through a different equation than in the \Planck\ cluster counts analysis \citep{PlanckXX1303.5080, PlanckXXIV1502.01597}, we show in Appendix~\ref{sec:connection} that it is the same mass bias parameter. In fact, we view our parametrization as a formally more correct way of defining the bias parameter $(1-\bsz)$, because it clearly identifies the connection of this parameter to the data within the context of a generative model. Our results will therefore be of direct relevance to the cluster cosmology analysis presented by \Planck.

Similarly, the parameter $(1 - \bl)$ characterizes any potential systematic bias in the CLASH lensing masses.  Any such bias would depend on the method used to extract the lensing masses, as well as specifics of the observations themselves, and can only be accurately estimated through survey-specific numerical simulations.  Generic simulations \mbox{\citep{Meneghetti2010a, Becker2011}} suggest that lensing masses for rich cluster systems, such as those in our sample, are unbiased at the few percent level, while \citet{Rasia2012} report underestimates of up to 10\%.  \citet{Meneghetti2014} simulated the CLASH sample in detail and concluded that lensing masses are unbiased, with $\sim (10-15)\%$ scatter, although they did not simulate the complete strong+weak lensing measurement analysis.  These studies provide a general idea of the level of possible bias and scatter, and we expect that in the near future simulations will improve the determination of these parameters and the slope of the mass dependence.     

The probability of a CLASH cluster having data $(\Mpl,\Mcl, \zspec)$ is
$$P(\Mpl, \Mcl \vert \Mfive, \bp) P(\zspec \vert z) d\Mpl d\Mcl d\zspec,$$
where
\bea\label{eq:mass_dist}
P(\Mpl, \Mcl \vert \Mfive, \bp) & = & \int d\ln\Msz d\ln\Ml \ P(\Mpl \vert \Msz)  \nn \\ 
& & \hspace{-1cm} \times P(\Mcl \vert \Ml) P(\ln\Msz, \ln\Ml \vert \Mfive, \bp),
\eea
$\bp$ is the vector of scaling relation parameters ($\bsz, \asz, \sigsz, \bl, \al, \sigl, r$) and $P(\zspec \vert z)$ is a delta function centered at $\zspec$. We consider the true mass and redshift of a cluster ($\Mfive, z$) as nuisance parameters.  The posterior probability distribution of our model parameters is then
\be\label{eq:posterior}
\lkhd(\bM, \bz, \bp \vert \vec{d}) \propto P_0(\bM, \bz \vert \bp) \prod_i P(\Mpl^{(i)}, \Mcl^{(i)} \vert \mu, \bp) P(\zspec^{(i)} \vert z),
\ee
where the product is over all galaxy clusters, the vector $\bM$ comprises the true cluster masses, $\vec{d}$ the data ($\vec{\Mpl}, \vec{\Mcl}, \vec{\zspec}$), and we have dropped a normalization constant (the marginal probability of the data) that depends only on $\vec{d}$. We adopted  the priors listed in Table~\ref{tab:priors} on our scaling relation parameters, leaving only $P_0(\bM, \bz)$, the prior on the mass and redshift vectors, $\bM$ and $\bz$, respectively.

The prior $P_0(\bM, \bz)$ depends on the expected mass and redshift distribution of CLASH-detected clusters.  Let $\ndet(\Mfive, z)$ be this distribution. Assuming CLASH selects clusters in the redshift range $\zmin$ to $\zmax$, the probability distribution of finding a cluster with mass $M^{(i)}$ and redshift $z^{(i)} \in [\zmin, \zmax]$  is  
\be
P(M^{(i)}, z^{(i)}) \ dM dz = \frac{\ndet(M^{(i)}, z^{(i)})}{\mathcal{N}}\ dM dz,
\ee
where $\mathcal{N}$ is the normalization factor defined by
\be
\int_{\zmin}^{\zmax} dz \int_0^\infty dM \, P(M, z) = 1, 
\ee
which yields
\be
\mathcal{N} = N_{\rm det} = \int_{\zmin}^{\zmax} dz \int_0^\infty dM \, \ndet(M, z).
\ee
Thus, we obtain that the prior is
\be
P_0(\bM, \bz) =  \prod_i \frac{\ndet(\Mfive^{(i)}, z^{(i)})}{N_{\text{det}}},
\ee
and our posterior becomes
\be
\lkhd(\bM, \bz; \bp \vert \vec{d}) = \prod_i \frac{\ndet(\Mfive, z)}{N_{\text{det}}} P(\Mpl^{(i)}, \Mcl^{(i)} \vert \Mfive, \bp) P(\zspec^{(i)} \vert z).
\ee

We marginalized over our nuisance parameters by integrating over the vectors $\bM$ and $\bz$ to obtain 
\be\label{eq:post}
\lkhd(\bp \vert \vec{d}) = \prod_i \lkhd_i (\bp \vert \vec{d}),
\ee
where we defined
\bea\label{eq:post_i}
\lkhd_i (\bp \vert \vec{d}) \equiv & \frac{1}{N_{\text{det}}} \int_{\zmin}^{\zmax} dz \int_{0}^\infty d\Mfive \ \ndet(\Mfive, z) \nonumber \\
& \times \ P(\Mpl^{(i)}, \Mcl^{(i)} \vert \Mfive, \bp) P(\zspec^{(i)} \vert z).
\eea
Given that the redshift distribution is a delta function, Eq.~\eqref{eq:post_i} reduces to
\be\label{eq:post_i2}
\lkhd_i  \equiv \frac{1}{N_{\text{det}}} \int_{-\infty}^\infty d\ln\Mfive \ \ndet(\Mfive, \zspec^{(i)}) P(\Mpl^{(i)}, \Mcl^{(i)} \vert \Mfive, \bp),
\ee
where $\zmin \leq \zspec^{(i)} \leq \zmax$.  Together with Eq.~\eqref{eq:post}, this is our final expression for the posterior distribution over the parameter space. It is worth noting that this posterior is equivalent to the one built taking into account the Poisson distribution of the missing clusters and then marginalizing over the number of these undetected objects (see, e.g., \cite{Mantz2010}).

We then modeled the selection function, that is, the mass-redshift distribution of the CLASH sample. Despite the detailed study by \cite{Meneghetti2014} quantifying the effects of the CLASH selection criteria on the determination of the concentration-mass relation, it is difficult to extract a precise selection function in terms of cluster mass from their simulations.  We therefore adopted the following approach.  
  
We assumed that CLASH selection in mass is redshift independent over the range $[\zmin, \zmax]$, and that the probability of a cluster being included in the sample is a function that goes to zero at low mass and to unity at high mass.  We modeled this with an error function:
\be\label{eq:sel_func}
f(\Mfive) = \frac{1}{2} \left[ 1 + \text{erf}\left( \frac{\ln \Mfive - \ln \Mcut}{\sqrt{2} \sigerf} \right) \right],
\ee
and treated the low-mass cutoff, $\Mcut$, and $\sigerf$, $\zmin$ and $\zmax$ as free parameters to be determined by the data themselves.  

We took the mass-redshift distribution as
\be\label{eq:M_z_dist}
\ndet (\Mfive, z) = f(\Mfive) \frac{dn(\Mfive, z)}{d\ln\Mfive} \frac{d^2V}{dzd\Omega},
\ee
where $dn/d\ln\Mfive$ is the halo mass function and $d^2V/dzd\Omega$ is the comoving volume element per unit solid angle, $d\Omega$. Throughout this paper, we have adopted the \citet{Tinker2008} multiplicity function with $\Mfive$ defined in terms of the critical density, and the $\LCDM$ cosmological parameters used to estimate both \Planck\ and CLASH masses as listed in Table~\ref{tab:cosmo_params}. 

\begin{table}
\begingroup
\newdimen\tblskip \tblskip=5pt
\caption{Fiducial cosmological parameters}
  \label{tab:cosmo_params}
\nointerlineskip
\vskip -3mm
\footnotesize
\setbox\tablebox=\vbox{
   \newdimen\digitwidth 
   \setbox0=\hbox{\rm 0} 
   \digitwidth=\wd0 
   \catcode`*=\active 
   \def*{\kern\digitwidth}
   \newdimen\signwidth 
   \setbox0=\hbox{+} 
   \signwidth=\wd0 
   \catcode`!=\active 
   \def!{\kern\signwidth}
\halign{#\hfil\tabskip=2em & \hfil#\tabskip=2em & \hfil#\tabskip=0pt\cr           
\noalign{\doubleline}
Parameter & Value \cr
\noalign{\vskip 3pt\hrule\vskip 1pt}
$H_0$ (km s$^{-1}$Mpc$^{-1}$) & $70.0$ \cr
$\Omega_b$ & $0.049$ \cr
$\Omega_m$ & $0.3$ \cr
$\Omega_{\Lambda}$ & $0.7$ \cr
$\sigma_8$ & $0.816$ \cr
$n_s$ & $0.967$ \cr
\noalign{\vskip 1pt\hrule\vskip 1pt}}}
\endPlancktable                    
\endgroup
\end{table}                        

\begin{figure*}
\centering
\includegraphics[width=1.05\textwidth]{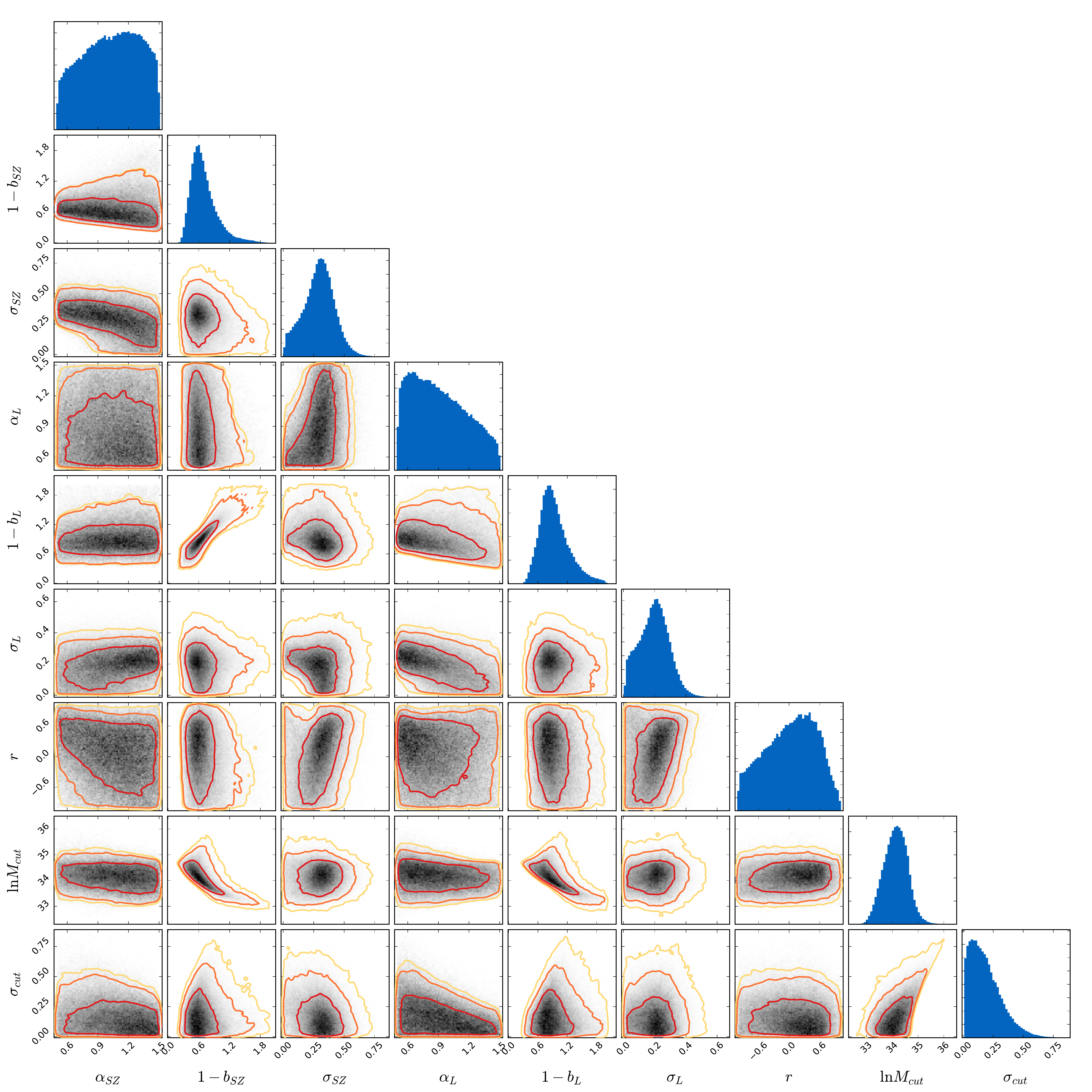}
\caption{Results of the full-parameter MCMC analysis with $1.4\times 10^{6}$ points (Case 1, Sect.~\ref{sec:res_all_free}). The contours correspond to the two-dimensional $68.3\%$, $95.4\%$, $99.7\%$ confidence regions on parameter pairs after marginalizing over the other parameters and the true mass and redshift [see Eqs.~\eqref{eq:post_i2}, \eqref{eq:sel_func} and \eqref{eq:M_z_dist}].  The histograms show the one-dimensional posterior marginal distributions for each parameter.\\}
\label{fig:esmcmc_1}
\end{figure*}



\subsection{Priors}
Our complete set of model parameters is
\be
\bp = \{\asz,\bsz,\sigsz,\al,\bl,\sigl,r,\Mcut,\sigerf,\zmin,\zmax\},
\ee
comprising four scaling relation parameters for the mean mass values ($\asz,\bsz,\al,\bl$), 
three scatter parameters ($\sigsz,\sigl,r$), and four selection function parameters ($\Mcut,\sigerf,\zmin,\zmax$).  The primary parameter of interest in this work is the bias, $\bsz$, of the SZ mass proxy and for which we adopted a flat prior.  The parameter $\bl$ is the possible bias in the CLASH lensing masses, for which we adopt the same flat prior as for $\bsz$ or a Gaussian prior with zero mean and standard deviation equal to 0.08 \citep{Umetsu2014}; we also performed analyses with fixed values of $\al=1$, $\bl = 0.0$, that is, unbiased lensing masses, and $\asz = 1$.  Table~\ref{tab:priors} summarizes the weak priors adopted on all other parameters.

\begin{table}                
\begingroup
\newdimen\tblskip \tblskip=5pt
\caption{Summary of priors}
  \label{tab:priors}
\nointerlineskip
\vskip -3mm
\footnotesize
\setbox\tablebox=\vbox{
   \newdimen\digitwidth 
   \setbox0=\hbox{\rm 0} 
   \digitwidth=\wd0 
   \catcode`*=\active 
   \def*{\kern\digitwidth}
   \newdimen\signwidth 
   \setbox0=\hbox{+} 
   \signwidth=\wd0 
   \catcode`!=\active 
   \def!{\kern\signwidth}
\halign{#\hfil\tabskip=2em & \hfil#\tabskip=2em & \hfil#\tabskip=0pt\cr          
\noalign{\doubleline}
Parameter & Interval (flat) & Gaussian \cr
\noalign{\vskip 3pt\hrule\vskip 1pt}
$\asz$ & $[0.5, 1.5]$ & -- \cr
$1 - \bsz$ & $[0.001, 2.0]$ & -- \cr
$\sigsz$ & $[0.01, 1.0]$ & -- \cr
$\al$ & $[0.5, 1.5]$ & -- \cr
$1 - \bl$ & $[0.001, 2.0]$ & $\text{G}(1.0, 0.08)$ \cr
$\sigl$ & $[0.01, 1.0]$ & -- \cr
$r$ & $[-1.0, 1.0]$ & -- \cr
\hline \cr
$\Mcut (h^{-1}\msun)$ & $[10^{12}, 10^{16}]$ & -- \cr
$\sigerf$ & $[0.01, 0.9]$ & -- \cr
$\zmin$ & $[0.0, 0.188]$ & -- \cr
$\zmax$ & $[0.89, 2.188]$ & -- \cr
\noalign{\vskip 1pt\hrule\vskip 1pt}}}
\endPlancktable                    
\endgroup
\end{table}                        

\subsection{Algorithm}
The posterior function is implemented in the Numerical Cosmology library \citep[NumCosmo,][]{DiasPintoVitenti2014}. The model and data objects are named \textsf{NcClusterPseudoCounts} 
and \textsf{NcDataClusterPseudoCounts}, respectively. The latter implements the $-2\ln\lkhd$ function and the former comprises, among other functions, $f(\Mfive)$, $\ndet$, $N_{\rm det}$ and the integral of Eq.~\eqref{eq:post_i2}.  In order to optimize computation time, we numerically calculate the three-dimensional integral over $\ln\Mfive$, $\ln\Msz$ and $\ln\Ml$ using the Divonne algorithm from the Cuba library \footnote{http://www.feynarts.de/cuba/}.  

The Gaussian probability distributions for the \Planck\ and CLASH masses, and the bivariate log-normal distribution for $\Msz$ and $\Ml$ (Eq.~\ref{eq:mass_dist}) are written in the \textsf{NcClusterMassPlCL} object. A detailed description of the mass function calculation is presented in \citet{Penna-Lima2014}. The python script (mass\_calibration\_planck\_clash.py) to reproduce the analyses presented in this work is distributed and available with NumCosmo\footnote{https://github.com/NumCosmo/NumCosmo}.  As an illustration, our code takes about 22 hours to carry out a Markov chain Monte Carlo (MCMC) study with $10^6$ points, using 100 chains and 40 cores.

\begin{table*}
\caption{Results -- The mean and $68\%$ CI of the marginal posterior distributions}             
\label{table:mean}      
\centering             
\begingroup
\setlength{\tabcolsep}{8pt} 
\renewcommand{\arraystretch}{1.5} 
\resizebox{\textwidth}{!}{
\begin{tabular}{c | c | c | c | c | c | c  }  
\hline\hline                 

\vtop{\hbox{\strut Fixed parameters}\vspace{0.2cm}\hbox{\hspace{0.3cm}\strut Prior on $\bl$}} & \multicolumn{1}{c |}{\vtop{\hbox{\strut None}\vspace{0.2cm}\hbox{\strut Flat}}} & \multicolumn{1}{c |}{\vtop{\hbox{\strut None}\vspace{0.2cm}\hbox{\strut Gaussian}}} & \multicolumn{1}{c |}{\vtop{\hbox{\strut None (without CLJ1226)}\vspace{0.2cm}\hbox{\hspace{0.9cm} \strut Gaussian}}} & \multicolumn{1}{c |}{\vtop{\hbox{\strut $\al = 1.0$}\vspace{0.2cm}\hbox{\strut Gaussian}}} & \multicolumn{1}{c |}{\vtop{\hbox{\strut $\asz = \al = 1.0$}\vspace{0.2cm}\hbox{\hspace{0.3cm} \strut Gaussian}}} &  \multicolumn{1}{c }{\vtop{\hbox{\strut $\asz = \al = 1.0$, $\bl = 0.0$}\vspace{0.2cm}\hbox{\hspace{1.3cm} \strut --}}} \\
& Mean $\pm 1\sigma$ & Mean $\pm 1\sigma$ & Mean $\pm 1\sigma$ & Median $\pm 1\sigma$ & Mean $\pm 1\sigma$ & Mean $\pm 1\sigma$ \\    
\hline                       
$\asz$ & $1.03^{+0.29}_{-0.33}$ & $1.05^{+0.29}_{-0.37}$ & $1.06^{+0.29}_{-0.35}$ & $1.04^{+0.31}_{-0.36}$ & -- & -- \\
$1 - \bsz$ & $0.71^{+0.45}_{-0.19}$ & $0.73^{+0.10}_{-0.09} $ & $0.78^{+0.11}_{-0.09}$ & $0.73^{+0.10}_{-0.08}$ & $0.73^{+0.09}_{-0.08}$ & $0.74 \pm 0.07$ \\
$\sigsz$ & $0.29^{+0.11}_{-0.14}$ & $0.28^{+0.12}_{-0.16}$ & $0.26^{+0.12}_{-0.13}$ & $0.31^{+0.11}_{-0.12}$ & $0.32^{+0.12}_{-0.09}$ & $0.32^{+0.11}_{-0.09}$ \\
$\al$ & $0.91^{+0.36}_{-0.26}$ & $0.87^{+0.39}_{-0.25}$ & $0.81^{+0.33}_{-0.19}$ & -- & -- & -- \\
$1 - \bl$ & $0.95^{+0.43}_{-0.23}$ & $0.996^{+0.076}_{-0.073}$ & $1.001 \pm 0.07$ & $0.993 \pm 0.08$ & $0.991^{+0.08}_{-0.07}$ & -- \\
$\sigl$ & $0.20^{+0.09}_{-0.11}$ & $0.20^{+0.08}_{-0.11}$ & $0.13^{+0.09}_{-0.07}$ & $0.17 \pm 0.10$ & $0.16^{+0.11}_{-0.09}$ & $0.17^{+0.10}_{-0.09}$ \\
$r$ & $0.03^{+0.49}_{-0.63}$ & $0.01^{+0.49}_{-0.66}$ & $0.16^{+0.51}_{-0.84}$ & $-0.04^{+0.54}_{-0.60}$ & $-0.07^{+0.53}_{-0.61}$ & $-0.05^{+0.47}_{-0.56}$ \\
\hline               
$\ln\Mcut$ & $34.13^{+0.44}_{-0.49}$ & $33.95^{+0.25}_{-0.17}$ & $33.93^{+0.30}_{-0.18}$ & $33.93^{+0.23}_{-0.15}$ & $33.94^{+0.23}_{-0.16}$ & $33.98^{+0.57}_{-0.15}$ \\
$\sigerf$ & $0.19^{+0.20}_{-0.11}$ & $0.17^{+0.14}_{-0.10}$ & $0.19^{+0.15}_{-0.11}$ & $0.15^{+0.12}_{-0.09}$ & $0.16^{+0.12}_{-0.10}$ & $0.18^{+0.14}_{-0.10}$ \\
$\zmin$ & $0.14^{+0.04}_{-0.12}$ & $0.13^{+0.04}_{-0.11}$ & $0.13^{+0.04}_{-0.11}$ & $0.13^{+0.04}_{-0.11}$ & $0.13^{+0.04}_{-0.11}$ & $0.13^{+0.04}_{-0.11}$ \\
$\zmax$ & $1.48^{+0.48}_{-0.41}$ & $1.50^{+0.44}_{-0.42}$ & $1.08^{+1.11}_{-0.28}$ & $1.50^{+0.48}_{-0.42}$ &  $1.50^{+0.49}_{-0.42}$ & $1.48^{+0.46}_{-0.40}$ \\
\hline                  
\end{tabular}}
\endgroup
\end{table*}

\section{Results}
\label{sec:results}
We explored six cases with the Bayesian analysis and present the results in Table~\ref{table:mean} and in Figs.~\ref{fig:esmcmc_1} and \ref{fig:esmcmc_2}. In the first full-parameter case, we leave all 11 parameters free and consider a flat prior on $\bl$ to understand the degeneracies inherent in the system.  As there is nothing to tie-down the overall mass scale of the sample, degeneracies appear between the mass bias parameter, $\bsz$, the lensing mass calibration, $\bl$, and the mass cut of the selection function.  In our second study, we also perform a full-parameter analysis, but now applying the Gaussian prior on $\bl$ in order to evaluate the effect of the lensing systematics on the determination of $\bsz$. We then examined three other cases by progressively adding strong constraints on the slopes of the lensing and SZ relations and also on $\bl$, namely: (i) $\al = 1.0$, (ii) $\asz = \al = 1.0$, (iii) $\asz = \al = 1.0$ and $\bl = 0.0$. Overall, fixing these parameters changes little on the constraints of the others, as shown in Table~\ref{table:mean}. 

Finally, we carried out a full-parameter study excluding the cluster CLJ1226 at $z = 0.89$. We observe an impact on SZ and lensing biases and scatters,  but these results are consistent with the constraints of the other five cases within the $68.3\%$ confidence interval (CI). 

\subsection{Case 1: All parameters free, flat prior on $\bl$}
\label{sec:res_all_free}
For our first study, we computed the joint posterior distribution for the full parameter set describing the SZ- and lensing-mass distribution and the selection function.  We ran 100 chains using the \textsf{NcmFitESMCMC} algorithm, an ensemble sampler with affine invariance for MCMC analysis from NumCosmo, requiring convergence of the variance of the fit parameters and $-2\ln\lkhd$, and of the multivariate potential scale reduction factor (MPSRF). The latter should be at least smaller than $1.2$, and $\text{Var}(-2\ln\lkhd)$ should be close to 22, since we are fitting 11 parameters. 

We computed a total of $1.5 \times 10^{6}$ sampling points and considered a burn-in of $10^{5}$ points, obtaining ${\rm MPSRF} \simeq 1.04$ and $\text{Var}(-2\ln\lkhd) \simeq 19.5$.  As a consistency check, we also calculated these values for different burn-in sizes, namely 10 equally spaced points between $[10^4, 10^5]$, confirming the convergence status of the chains.  Figure~\ref{fig:esmcmc_1} shows the $68.3\%$, $95.4\%$ and $99.7\%$ confidence regions for parameter pairs, as well as the one-dimensional marginal distribution for each parameter.  

It is worth noting that, in high-dimensional parameter space MCMC does not provide, in general, accurate estimates for the best fit\footnote{For instance, consider a n-dimensional unit Gaussian, whose maximum is at the origin. The probability of  the number of samples to be close to the maximum is small, since the volume of a high-dimensional sphere is concentrated in a narrow annulus at the surface \citep{Unpingco2016}.}  \citep{Lewis2002,Hobson2009}. Therefore, in Table 4 we quote the mean and the $68.3\%$ CI of the marginal posterior of each parameter. We determined the $68.3\%$ CI of the i-th parameter $p_i$ by finding the points $p_i^-$ and $p_i^+$ such that the probability $\text{Pr}(p_i^- \leqslant p_i \leqslant \bar{p}_i) = 34.15\%$ (68.27\%/2) and $\text{Pr}(\bar{p}_i \leqslant p_i \leqslant p_i^+) = 34.15\%$, respectively, where $\bar{p_i}$ is the mean of $p_i$. 

In general, the marginal  distributions are highly non-Gaussian. The parameters are very degenerate in this first, unconstrained exploration, most notably the slopes $\asz$ and $\al$, the correlation $r$ and $\zmax$.  We see in Fig.~\ref{fig:esmcmc_1} that their confidence regions cover the entire range of values defined by their flat priors, and that their errors are of the order of $50\%$ and larger (see Table~\ref{table:mean}, column ``none, flat'').

The mass bias parameter, $(1-\bsz)$, is strongly correlated with the lensing mass calibration, $(1-\bl)$, and both are strongly anti-correlated with $\Mcut$.  The former correlation is easily understood, because the mass bias is obviously tied to halo mass through the lensing measurements; changing the lensing calibration correspondingly changes the mass bias parameter.  The anti-correlation between $\bl$ and $\Mcut$ is a result of the lack of any absolute mass tie-down in this full parameter exploration: the system is attempting to calibrate the overall mass scale, and hence the lensing mass bias, through the selection function mass cut-off.  This anti-correlation then spills into $\bsz$ through its correlation with $\bl$.

The SZ and lensing scatters are reasonably well constrained by the data, with large values, $\sigsz \gtrsim 0.6$ and $\sigl \gtrsim 0.4$, disfavored.  At $0.20$, the mean for $\sigl$ is consistent with expectations based on the simulations by \cite{Meneghetti2014}. We also see from Fig.~\ref{fig:esmcmc_1} that $\sigsz$ ($\sigl$) is moderately anti-correlated with $\asz$ ($\al$). We note this anti-correlation refers
to the uncertainties in $\sigsz$ and $\sigl$, and not to the correlation coefficient between these two, which is unconstrained by the data.


\subsection{Case 2: All parameters free, Gaussian prior on $\bl$}
\label{sec:case2}
\begin{figure*}
\centering
\includegraphics[width=1.05\textwidth]{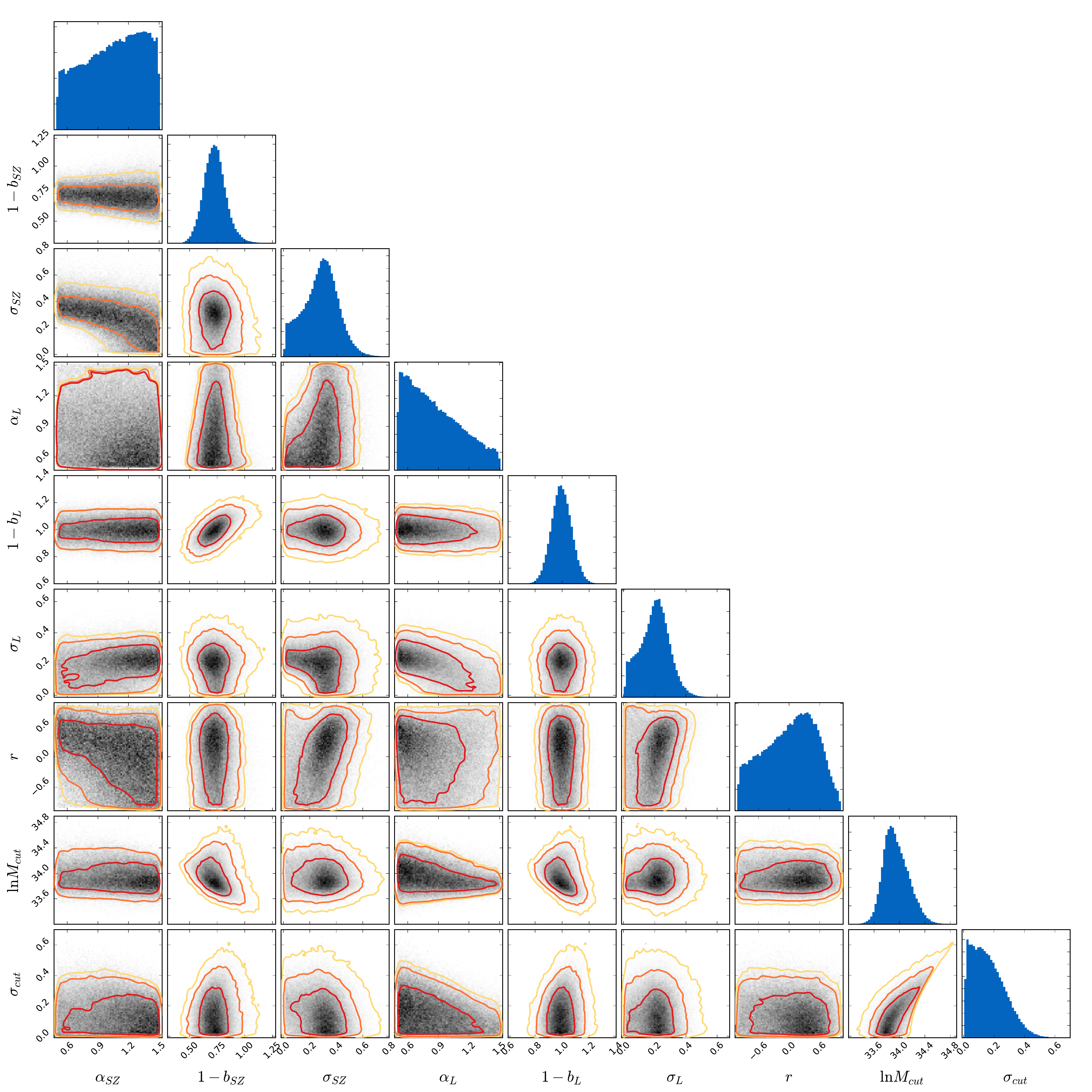}
\caption{Results of the full-parameter MCMC analysis assuming a Gaussian prior of $\bl=0\pm 0.08$, with $8.5\times 10^{5}$ points (Case 2, Sect.~\ref{sec:case2}) and following the format of Fig.~\ref{fig:esmcmc_1}.\\}
\label{fig:esmcmc_2}
\end{figure*}

Similarly to the previous study, we now fit the 11 parameters considering a Gaussian prior on $\bl=0.0 \pm 0.08$ \citep{Umetsu2014, Meneghetti2014}.  We performed an MCMC analysis generating $8.5 \times 10^{5}$ sample points. With a burn-in size of $10^{5}$, we obtained ${\rm MPSRF} \simeq 1.03$ and $\text{Var}(-2\ln\lkhd) \simeq 15.9$. The marginal distributions and the $68.3\%$, $95.4\%$ and $99.7\%$ confidence regions are shown in Fig.~\ref{fig:esmcmc_2}.  Given the lack of correlation between $\asz$, $\sigsz$, $\al$, $\sigl$, $r$, $\sigerf$, $\zmin$ and $\zmax$ with $\bl$, there is no improvement in their constraints (see third column of Table~\ref{table:mean}). 

The main differences concern $\bsz$ and $\Mcut$, which were strongly correlated with $\bl$ in Case 1.  We drastically reduce the uncertainty on $(1-\bsz)$, by about $70\%$ with mean $0.73$ and $68.3\%$ CI of $[0.64, 0.83]$.  Similarly, for $\ln\Mcut$ the decrease in the $68.3\%$ CI is $\sim60\%$.  In addition, fixing the mass scale tightens the correlation between $\ln \Mcut$ and $\sigerf$.  

Another effect of the $\bl$ prior is to weaken the correlation between $\bsz$ and $\Mcut$, confirming that their previous strong correlation leaks through from their relation to $\bl$.  This means that our constraints on the mass bias, $\bsz$, are relatively insensitive to the selection function as long as the lensing measurements robustly tie-down the mass scale.

\subsubsection{Other cases}
\label{sec:other_cases}
We now consider more specific cases by fixing (i) $\al = 1.0$, (ii) $\asz = \al = 1.0$, and (iii) $\asz = \al = 1.0$ and $\bl = 0.0$ in Eqs.~\eqref{eq:avgmsz}--\eqref{eq:avgml}. This last corresponds the case of calibrating the overall scale of the \Planck\ mass proxy assuming the CLASH masses are unbiased. We ran 100 chains, computing $5 \times 10^5$ points with a burn-in of $10^5$ for each case. The respective results are given in the fifth, sixth and seventh columns of Table~\ref{table:mean}. 

All three cases indicate that the constraints are limited by the small statistics of the sample. In general, the parameters are not correlated and, therefore, fixing the SZ and lensing slopes of the mass scaling relations do not tighten the constraints on the remaining parameters. The main difference is the reduction in  $\sim 30\%$ of the $68.3\%$ CI of $(1 - \bsz)$, when considering the extreme case (iii), in which there are no systematics in the lensing masses. In fact, the uncertainty on $(1-\bsz)$ in Case 2 is the quadrature sum of the statistical uncertainty here and the Gaussian uncertainty on $\bl$: $0.1=\sqrt{0.07^2+0.08^2}$, as could be expected.  It is worth mentioning that the regularity of these three $(1 - \bsz)$ estimates lends support to the conservative result presented in Sect.~\ref{sec:case2}.

We also fit the 11 parameters removing the outlier CLJ1226 from the cluster catalog. Running also 100 chains, we generated $5 \times 10^5$ points to reach ${\rm MPSRF} \simeq 1.09$ and $\text{Var}(-2\ln\lkhd) \simeq 22.06$. From Fig.~\ref{fig:data}, we would expect an increase in the correlation, a decrease in the SZ and lensing scatters and in both bias parameters, since the outlier seems to require a spreader distribution to include it. The results, displayed in the fourth column of Table~\ref{table:mean}, confirm these expectancies. For instance, $1-\bsz = 0.78$, $\sigma_l = 0.13$ and $r = 0.16$, although they are not statistically significant. We note that the results of all six cases are consistent, in part due to the broad constraints on the parameters, and that future applications of our methodology to larger catalogs promise to break the degeneracies.
    
\subsection{Non-Gaussian prior}
\label{sec:case7}

The results presented so far show that the constraints on $\bsz$ are strongly dependent of $\bl$. In addition to the flat and Gaussian priors on $\bl$, and the unbiased case ($\bl = 0.0$), we now consider one last case study with a non-Gaussian prior.

\begin{figure}
        \centering
        \includegraphics[width=.5\textwidth]{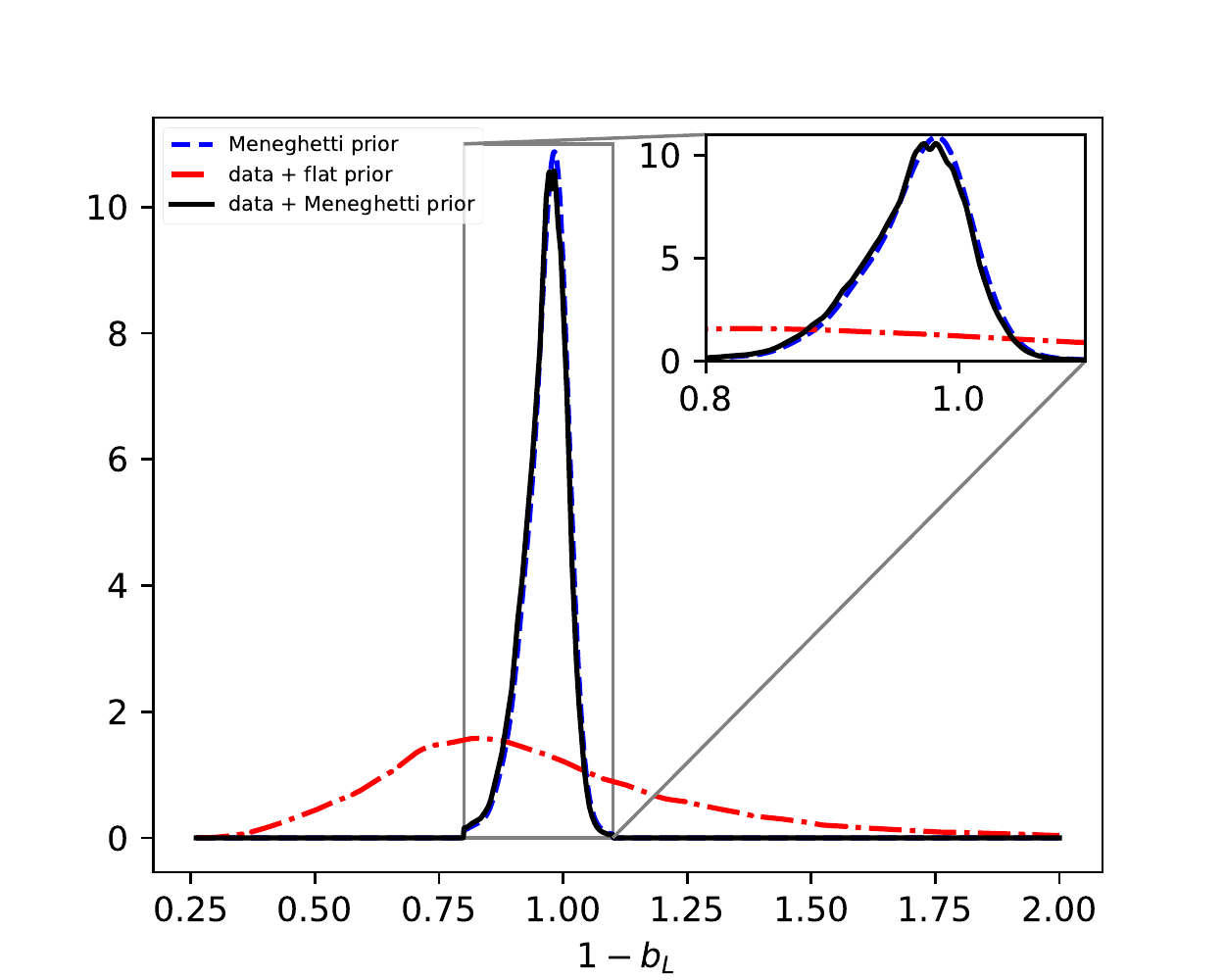}
        \caption{Prior distributions for the lensing mass bias parameter, $\bl$.  The blue dashed line represents the Meneghetti prior, i.e., the 3D lensing-true mass ratio distribution \citep{Meneghetti2014}. The other curves give the $1-\bl$ posterior distributions in the case of flat (red line) and Meneguetti (black line) priors obtained from the MCMC analyses of the CLASH-\Planck\ cluster sample (data).}  
        \label{fig:bl_meneghetti}
\end{figure}

This new prior is based on the result presented in \citet{Meneghetti2014}. The authors obtained the distribution of the 3D lensing-true mass ratio considering the NFW profile (among others). This is a left-skewed distribution in the interval $(1-b_L) \in [0.8, 1.1]$, which we use as a prior on $b_L$ and show in Fig.~\ref{fig:bl_meneghetti} (blue dashed line) labeled as the Meneghetti prior.

\begin{figure}
        \centering
        \includegraphics[width=.5\textwidth]{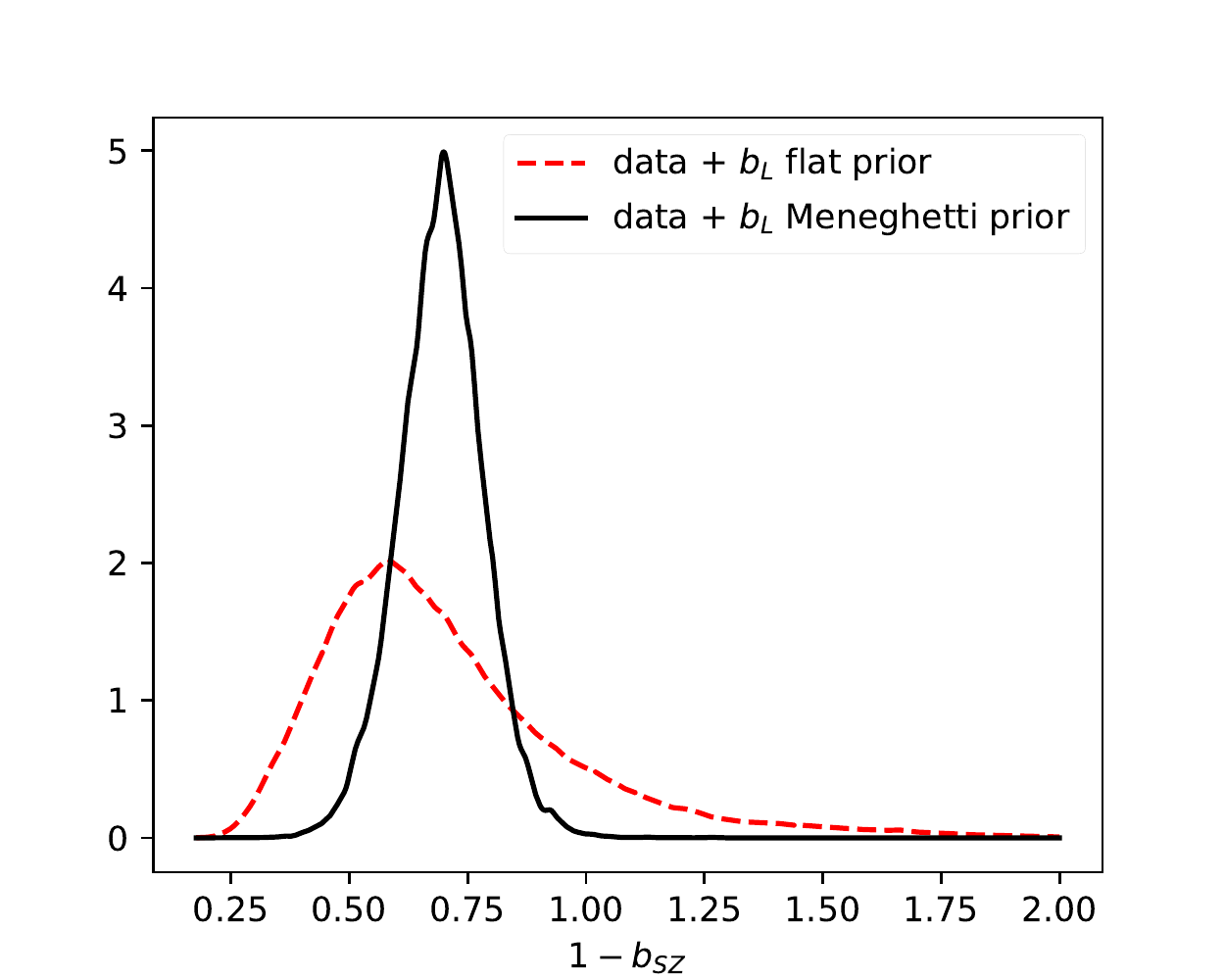}
        \caption{The posterior distribution of $1-\bsz$ for the flat (red line) and Meneguetti (black line) priors.}  
        \label{fig:bsz_meneghetti}
\end{figure}

In this case we ran 100 chains, computing $6\times 10^5$ sampling points (burn-in size of $10^5$). The posterior distributions of $(1-\bl)$ and $(1-\bsz)$ are shown in Figs.~\ref{fig:bl_meneghetti} and \ref{fig:bsz_meneghetti}, respectively. We compared them with the posteriors obtained from the MCMC analysis considering the flat prior. Similar to the previous cases, we see the strong effect on $(1-\bsz)$ due to the prior on $(1-\bl)$.

Table~\ref{table:mean_meneghetti} displays the mean and the 68.3\% CI of all 11 parameters for the Gaussian and Meneghetti priors, second and third columns respectively. For instance, as the Meneghetti prior restricts $\bl$ to a narrower interval, naturally its error bar is accordingly reduced in comparison to the flat and Gaussian priors, namely, $(1-\bl) = 0.964_{-0.057}^{+0.034}$. As expected, given the form of the Meneghetti prior (blue dashed line in Fig.~\ref{fig:bl_meneghetti}), the lensing-mass proxy now presents a small bias. Consequently, the SZ bias increases by 3\% to $(1-\bsz) = 0.69_{-0.09}^{+0.08}$, in comparison to the Gaussian prior centered in $\bl = 0$, whereas the error bar decreases due to the narrower $\bl$ interval. It is worth mentioning that both results are in accordance within $68.3\%$ CI.

On the other hand, the parameters of the selection function, $\Mcut$ and $\sigerf$, increase in both their mean values and their uncertainties even when compared to the flat-prior case (see Tables~\ref{table:mean} and \ref{table:mean_meneghetti}). This is due to the anti-correlation between $\bl$ and $\Mcut$. The $(1-\bl)$ interval, [0.8, 1.1], favors larger values of $\Mcut$ and, consequently, $\sigerf$. The remaining parameters present no significant modification compared to the previous cases.


\begin{table}
        \caption{Results -- The mean and $68\%$ CI of the marginal posterior distributions}             
        \label{table:mean_meneghetti}      
        \centering             
        \begingroup
        \setlength{\tabcolsep}{8pt} 
        \renewcommand{\arraystretch}{1.5} 
        {
                \begin{tabular}{c | c | c }  
                        \hline\hline                 
                        
                        \vtop{\hbox{\strut Fixed Parameters}\vspace{0.2cm}\hbox{\hspace{0.3cm}\strut Prior on $\bl$}} & \multicolumn{1}{c |}{\vtop{\hbox{\strut None}\vspace{0.2cm}\hbox{\hspace{-0.3cm} \strut Gaussian}}} & \multicolumn{1}{c }{\vtop{\hbox{\hspace{0.3cm} \strut None }\vspace{0.2cm}\hbox{\strut Meneghetti}}} \\
                        & Mean $\pm 1\sigma$ & Mean $\pm 1\sigma$ \\    
                        \hline                       
                        $\asz$ & $1.05^{+0.29}_{-0.37}$ & $1.03^{+0.29}_{-0.32}$ \\
                        $1 - \bsz$ & $0.73^{+0.10}_{-0.09} $ & $0.69^{+0.08}_{-0.09}$ \\
                        $\sigsz$ & $0.28^{+0.12}_{-0.16}$ & $0.28^{+0.12}_{-0.15}$ \\
                        $\al$ & $0.87^{+0.39}_{-0.25}$ & $0.83^{+0.41}_{-0.21}$ \\
                        $1 - \bl$ & $0.996^{+0.076}_{-0.073}$ & $0.964^{+0.034}_{-0.057}$ \\
                        $\sigl$ & $0.20^{+0.08}_{-0.11}$ & $0.20^{+0.09}_{-0.11}$ \\
                        $r$ & $0.01^{+0.49}_{-0.66}$ & $0.04^{+0.48}_{-0.65}$ \\
                        \hline               
                        $\ln\Mcut$ & $33.95^{+0.25}_{-0.17}$ & $34.14^{+0.97}_{-0.24}$ \\
                        $\sigerf$ & $0.17^{+0.14}_{-0.10}$ & $0.22^{+0.21}_{-0.12}$ \\
                        $\zmin$ & $0.13^{+0.04}_{-0.11}$ & $0.13^{+0.04}_{-0.11}$ \\
                        $\zmax$ & $1.50^{+0.44}_{-0.42}$ & $1.50^{+0.44}_{-0.43}$ \\
                        \hline                  
        \end{tabular}}
        \endgroup
\end{table}
    

\clearpage
\pagebreak
\clearpage
\pagebreak


\section{Discussion}
\label{sec:discussion}
Our basic result is a constraint on the \Planck\ mass bias parameter of $(1-\bsz)=0.73\pm 0.10$ ($0.69_{-0.09}^{+0.08}$).  This value applies at the median mass of our sample, $M_0 = 5.7\times 10^{14}h^{-1}$, and our fit is consistent with no mass dependence, although with large uncertainty.  Obtained by fitting all 11 parameters and assuming the Gaussian (Meneghetti) prior on $\bl$ (Cases 2 and 7, Sects.~\ref{sec:case2} and \ref{sec:case7}, second and third columns Table~\ref{table:mean_meneghetti}, respectively), these results agree within the uncertainties with the constraints in the five other case studies, as well as with the sample mean ratio from Sect.~\ref{sec:mass_comparison}. They improve on the latter by folding in astrophysical effects, such as intrinsic, correlated scatter, and the influence of the CLASH selection function.  

We have used \Planck\ as a follow-up to the CLASH sample and, as a consequence, are not affected by the selection (Malmquist\footnote{Classical Malmquist bias applies to a flux limited sample and refers to the fact that intrinsically more luminous objects are over-represented because they can be seen over larger volumes than less luminous sources. In common practice, Malmquist bias is the term applied generally, but inaccurately, to effects related to sample selection.}) bias noted by \citet{Battaglia2016}\footnote{The authors referred to this correction as an Eddington bias correction.  Eddington bias is not a sample selection effect, but rather due to dispersion in an observable in the presence of a steep abundance function.  While discussing Eddington bias at an earlier point in their paper, the correction made to WtG and CCCP results due to missing clusters is more appropriately referred to as Malmquist bias.}.  This bias arises in the WtG and CCCP studies because some of their clusters do not have \Planck\ detections\footnote{We note, however, that we are unable to use one cluster in our CLASH sample because of its negative signal-to-noise value.}.  \citet{Battaglia2016} attempted to correct the WtG and CCCP results for this effect by assigning \Planck\ masses to the undetected clusters.  Without knowing which masses to assign, however, this correction is of course uncertain, as they discuss (see also below).  

Selection effects in our study would come through the CLASH sample definition.  Our Bayesian approach aims to fully account for any such effects by incorporating the sample selection function into the analysis, and will do so to the extent that our model of the selection function is accurate.  Ideally, we would like to know the sample selection function a priori.  The CLASH sample selection is complex, and \citet{Meneghetti2014} use detailed numerical simulations to evaluate its impact on determination of the mass-concentration relation for clusters.  Unfortunately, it is not straightforward to extract a selection function in terms of true halo mass from this work.  We have instead opted to parameterize the CLASH selection function with the expected cosmological mass function and a smooth, but generic cutoff.  By studying their relation to the other parameters, we conclude that the exact values for the selection function parameters do not significantly impact our final constraint on the SZ mass bias.  This statement, however, is only as good as the generic form that we have employed for the selection function.

Our result improves on previous cluster mass calibrations by explicitly accounting for sample selection and associated uncertainties.  Moreover, we consistently account for other statistical effects, such as Eddington bias, related to dispersion in the cluster observables and to their possible correlation.  This is clear from our likelihood expression (Eq.~\ref{eq:post_i2}), which is a convolution of the steep prior on true mass ($\Mfive$) with the multivariate cluster observable distribution.  

The Bayesian analysis also provides information on the intrinsic dispersion of the SZ proxy and lensing masses, and their correlation.  In all seven cases, we find a 13-20\% scatter for the lensing mass, in good agreement with expectations \citep{Meneghetti2014}.  The estimates also indicate a 30\% scatter for the SZ mass proxy, notwithstanding, it is not well constrained, and remains consistent with the $\sim 10\%$ scatter expected from simulations \citep[e.g.,][]{Nagai2007}. Unfortunately, we are unable to establish a meaningful constraint on the correlation, $r$. We note that our results for the intrinsic lensing and SZ scatter masses are compatible with, respectively, the lensing and hydrostatic equilibrium  scatters obtained by \citet{Sereno2015}. 

We find a value for the \Planck\ SZ mass bias that is consistent with the constraints $(1-\bsz)=0.688\pm 0.072$ and $(1-\bsz)=0.76\pm 0.05$ (stat) $\pm 0.06$ (syst) reported, respectively, by the WtG and CCCP lensing programs \citep{vonderLinden2014, Hoekstra2015}.
This agreement holds even after the correction to the WtG and CCCP values proposed by \citet{Battaglia2016}, apart from the most extreme cases.  This is satisfying because the samples and the
lensing mass extraction methodologies differ significantly. The wide-field ground-based data for 
WtG and CCCP is augmented in CLASH by deep, 16-band HST data for weak -and strong-lensing, which the 
{\tt SaWLens} reconstruction method combines into a single two-dimensional reconstruction. While doing so, it makes no assumption about the underlying mass distribution causing the lensing signal. This method differs to the aforementioned studies, which either rely on parametric fits or an aperture mass applied to the weak-lensing shear data only. The different mass estimates differ thus in both reconstruction methodology and input data.

The LoCuSS collaboration finds $(1 - \bsz) = 0.95 \pm 0.04$, based on their sample of 50 clusters at $0.15<z<0.3$ \citep{Smith2016}. They show that within the uncertainties their result is consistent with both
CCCP and WtG results discussed above, when the CCCP and WtG samples are restricted to clusters at $z<0.3$.  Moreover, the LoCuSS weak-lensing cluster masses for five clusters in the CLASH sample are in excellent agreement with CLASH measurements \citep{Okabe2016}. This implies that $1 - \bsz$ may evolve with redshift.  Specifically, Smith et al. found $(1 - \bsz) = 0.6 \pm 0.1$ for WtG and $(1 - \bsz) = 0.7 \pm 0.1$ for CCCP, both at $z > 0.3$. These measurements at $z > 0.3$ are fully consistent with our basic results of $(1 - \bsz) = 0.73 \pm 0.10$ ($0.69_{-0.09}^{+0.08}$), based on a sample that is dominated by clusters at $z > 0.3$.

The primary motivation for all these studies is to quantify the extent of the tension between constraints from the primary CMB and cluster counts found by \Planck.  The greatest source of uncertainty in this tension is  presently the \Planck\ cluster mass calibration.  In \cite{PlanckXX1303.5080, PlanckXXIV1502.01597}, the mass bias is defined through the SZ signal - halo mass relation, where no bias ($\bsz=0$) corresponds to the mass calibration based on \XMM\ X-ray observations, as detailed in the appendix of \citet{PlanckXX1303.5080}.  Our definition here is based on the \Planck\ SZ mass proxy, $\Mpl$, described in Sect.~\ref{sec:Mpl}.  While not immediately obvious, Appendix~\ref{sec:connection} demonstrates that the two are equivalent and that the mass bias constrained here is in fact the same as that used in the \Planck\ cluster cosmology analyses.  

Figure~\ref{fig:cosmology} summarizes the implications for the \Planck\ cluster cosmology results.  In it we assemble a number of recent mass calibration results and compare them to the value of $(1-\bsz)=0.58\pm 0.04$ required (yellow band) by the \Planck\ primary CMB cosmology, as deduced in \citet{PlanckXXIV1502.01597} when leaving the mass bias parameter free.  Measurements published before the 2013 \Planck\ cosmology results are also included in the figure.  These earlier studies do not report results in terms of the mass bias parameter, but rather the normalization of the $Y$-$M$ relation.  The mass bias parameter, $(1-\bsz)$, is a parameterization of this amplitude that became standard afterwards when referring to the \Planck\ cluster mass scale.  

In the figure, P11 refers to the work by \citet{P11} that defines the reference point where $\bsz=0$.    The values labeled M10 and V09 are obtained by rewriting the amplitude of the $Y$-$M$ relation of \citet{Rozo2014b} in terms of the $(1-\bsz)$ parameter.  These amplitudes were derived using a self-consistent method of propagating scaling relations that has since been improved upon by \citet{evrard2014}.  The point R09 is the predicted amplitude for $(1-\bsz)$ derived from the combined SZ and weak lensing signals of maxBCG galaxy clusters using the same methods of \citet{Rozo2014b}.  The R14 point corresponds to the preferred scaling relation of \citet{rozo2014c}, which combined the maxBCG and V09 $Y$-$M$ scaling relation, correcting the former downwards in mass by 10\%, and the latter upwards in mass by 21\%.  These two sets of scaling relations were then combined at the likelihood level to arrive at the R14 point.  The PXX point corresponds to \citet{PlanckXX1303.5080}, who took $(1-\bsz)=0.8$ as their fiducial value, but adopted a top-hat systematic error budget $(1-\bsz)\in [0.7,1]$, delineated here by the error bars.  

The WtG and CCCP points correspond to the work of \citet{vonderLinden2014} and \citet{Hoekstra2015}, who calibrated $(1-\bsz)$ based on a comparison of the SZ masses from \Planck\ with their weak lensing mass estimates using the subsample of \Planck\ clusters with weak lensing follow up from each of these groups.  \citet{Battaglia2016} noted that the incomplete overlap of the cluster samples introduces a bias in the recovered $(1-\bsz)$ parameter (Malmquist bias), and we indicate the suggested corrections from \citet{Battaglia2016} in Figure ~\ref{fig:cosmology}.  The uncertainty on this correction increases the error bars.  The Simet15 point comes from \citet{Simet2017}, based on a stacked weak lensing analysis of the MCXC cluster catalog.  Finally, the LoCUSS point is a $(1-b_{SZ})$ estimate from the LoCUSS collaboration \citep{Smith2016} that compares the Planck and lensing mass estimates. 

The two main results from this paper are the points labeled CLASH(G) and CLASH(M), referring to the Gaussian and Meneghetti priors.  Their error bars are larger than the comparable studies by WtG and CCCP because, in the Gaussian case, for instance, we have incorporated an 8\% uncertainty, centered at zero, on any potential bias in lensing mass estimates.  All of the mass calibrations lie above the range favored by the \Planck\ primary CMB cosmology, although almost none of the more recent values does so with notable significance on its own.  It is important to reduce the uncertainties in cluster mass calibration to obtain a more clear understanding of the existence of any tension between the cluster counts and the primary CMB constraints.

As a final note, we consider the effect of the lower reionization optical depth, $\tau$, reported by \citet{Adam2016} in an updated analysis of large angular scale polarization in \Planck.  The CMB determines the combination $A_{\rm s}e^{-2\tau}$ to high precision, where $A_{\rm s}$ is the power spectrum amplitude on large scales and is $\propto \sigma_8^2$, assuming all other cosmological parameters are fixed.  Lowering the optical depth therefore lowers $\sigma_8$ from the primary CMB and moves the yellow band in Fig.~\ref{fig:cosmology} upwards.  Taking the central value of $\tau=0.058$ given by \citet{Adam2016}, we estimate that the center line of the yellow band increases by $\sim 8$\% to $(1-\bsz) \approx 0.63$.


\section{Conclusion}
\label{sec:conclusion}
\begin{figure}
\centering
\includegraphics[width=.5\textwidth]{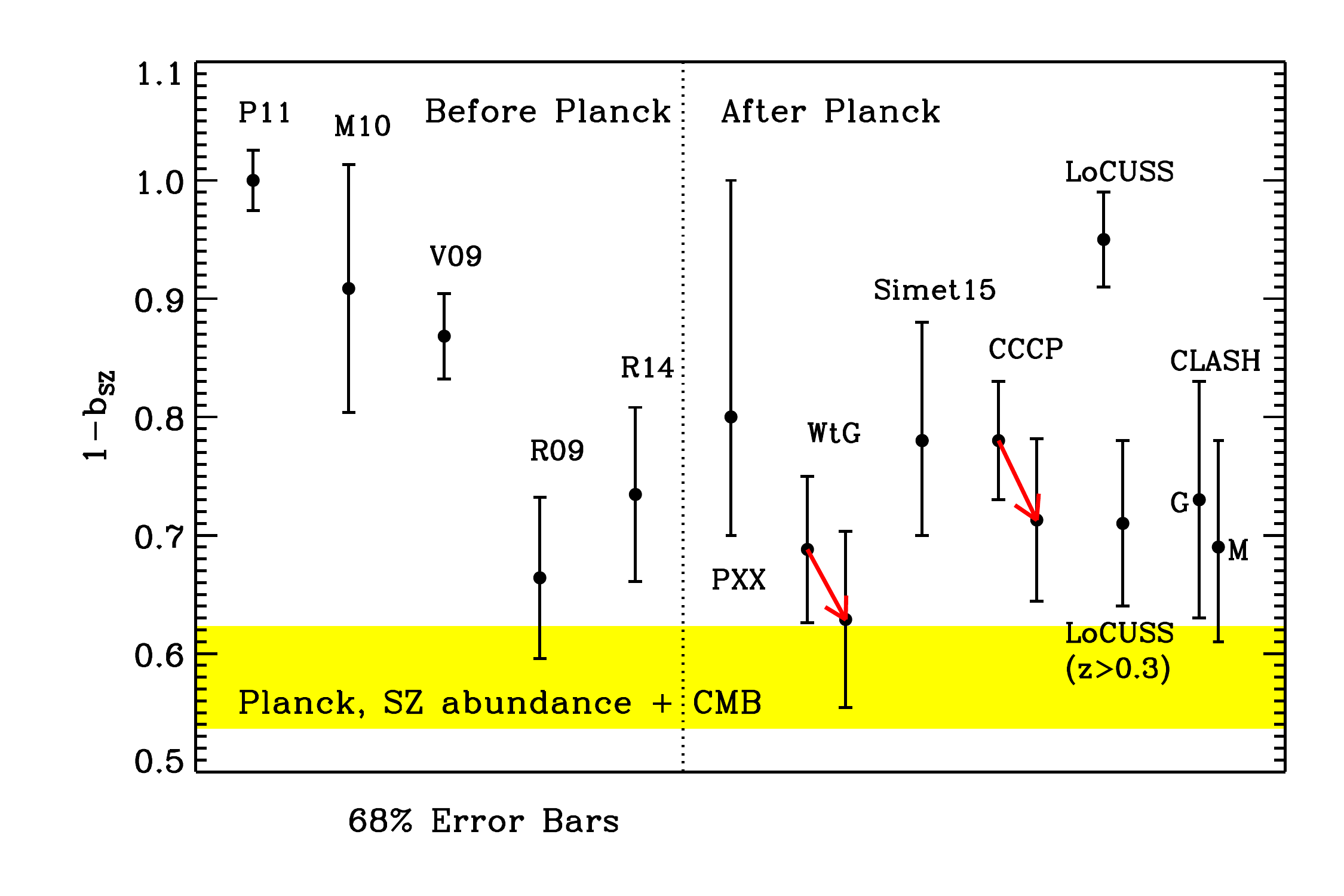}
\caption{Mass bias parameter, $(1-\bsz)$, measured in different studies compared to the values preferred by the \Planck\ base $\Lambda$CDM fit to the primary CMB anisotropies \citep{PlanckXXIV1502.01597}, shown as the yellow band. For reference, we translate results published before the \Planck\ 2013 cosmological constraints became available to values for $(1-\bsz)$. These are P11 \citep{Planck2011}, M10, V09, R09 \citep{Rozo2014b} and R14 \citep{rozo2014c}.  PXX is the 2013 SZ cluster cosmology analysis by \Planck\ \citep{PlanckXX1303.5080} and represents the top-hat range used as a prior in that analysis.  The WtG and CCCP points are, respectively, from \citet{vonderLinden2014} and \citet{Hoekstra2015}, with the red arrows showing the corrections estimated by \citet{Battaglia2016}; the error bars increase to include the uncertainty on this correction.  The \citet{Simet2017}, based on lensing stacks of the MCXC catalog, and the LoCUSS \citep{Smith2016} results are reproduced as labeled.  See text for details.}  
\label{fig:cosmology}
\end{figure}

We are in the process of gaining considerable insight into the cluster mass scale thanks to recent samples of tens of clusters with high quality lensing mass determinations, now reaching statistical constraints of $\sim 10\%$.  These constraints are fundamental to cluster cosmology.  Fig.~\ref{fig:cosmology} summarizes recent determinations of the \Planck\ cluster mass bias parameter and compares them to earlier mass bias estimates and to the value required by the \Planck\ primary CMB cosmology.  

The lensing based determinations (R09, R14, WtG, Simet, CCCP, CLASH) mostly display a coherent picture within the statistical uncertainties. While the value for $1-\bsz$ reported by LoCUSS is inconsistent with ours, LoCUSS sees a strong redshift evolution in $1-\bsz$, and their final $1-\bsz$ value is dominated by clusters that are lower redshift than the bulk of the CLASH sample.  Our result is consistent with LoCUSS for $z>0.3$, see Fig.~\ref{fig:cosmology}, as discussed in Sect.~3.2 of \citet{Smith2016}.  Malmquist and Eddington bias affect some of the points to an uncertain degree.  \citet{Battaglia2016} estimated the Malmquist bias corrections on the WtG and CCCP determinations, which we indicate in the figure.  Our constraint fully accounts for these, as well as astrophysical effects.

Other systematic effects may remain, however, at an important level.  For example, in our study we adopted a generic form for the CLASH selection function. It would be far better to have a form that is well motivated from simulations, something that continued examination of the \citet{Meneghetti2014} simulations could afford.  Similarly, detailed simulations are needed to evaluate the possible bias in lensing mass measurements (i.e., $\bl$ and $\al$).  They must reproduce both the sample selection and the specific lensing mass extraction methodology.  The technique to achieve such comprehensive simulations exists, but the studies have yet to be performed.  It is important to emphasize in this light that each cluster cosmology sample must be analyzed in its own specific context.

With these recent advances we have perhaps learned more about {\em how} to calibrate the mass scale than we have actually improved understanding of the tension between the \Planck\ primary CMB and cluster cosmology constraints.  All mass scale determinations lie high relative to the preferred CMB value, although in each case the significance is low. This is also true of our measurement.  In this context, it is important to note that the lower optical depth to reionization recently reported by  \citet{Adam2016} shifts the yellow band in Fig.~\ref{fig:cosmology} upwards from a center line of $(1-\bsz)=0.58$ to $(1-\bsz)\approx 0.63$, according to our approximate calculation.

Progress in understanding is encouraging and emphasizes the importance of improving constraints beyond the current level.  With such progress on relatively small samples, the future looks promising with large lensing programs like \Euclid\ \citep{Laureijs2011}, WFIRST \citep{Spergel2013} and the Large Synoptic Survey Telescope \citep{LSST2009} that will produce samples of thousands of objects thanks to their wide-field surveying.  In addition, CMB lensing, already observed over large sky areas by \Planck, SPT and ACT, adds a powerful and independent method for mass measurements \citep{melin2015, baxter2015, madhavacheril2015}.  The Bayesian methodology presented here will be important to extract the full potential of these large datasets. 


\begin{acknowledgements}
MPL was supported by National Council for Scientific and Technological Development - Brazil (CNPq grant
202131/2014-9 and PCI/MCTIC/CBPF program). MPL and JGB thank Sandro Vitenti for valuable discussions, and Graham Smith for suggestions. ER acknowledges support from DOE grant DE-SC0015975 and from the Sloan Foundation, grant FG-2016-6443. JM has received funding from the People Programme (Marie Curie Actions) of the European Unions Seventh Framework Programme (FP7/2007-2013) under REA grant agreement number 627288. 

Part of this research was carried out at the Jet Propulsion Laboratory, California Institute of Technology, under a contract with the National Aeronautics and Space Administration.
\end{acknowledgements}

\bibliographystyle{aa}
\bibliography{clashplanck}
 

\appendix

\section{Connection to the \Planck\ SZ Cosmology Analysis}
\label{sec:connection}

The main result of this work is a constraint on the relation between a halo's true mass $\Mfive$
and the mass proxy $\Msz$.  It is not necessarily obvious how this result is related to the parameter
$b$ used in PXX, a point that we discuss in some detail here.

Let us then define a parameter $\beta$ such that 
\be
\avg{\ln \Msz|\Mfive} = \ln (1-\beta) + \ln \Mfive.
\ee
We will show that $\beta=b$, where $b$ is defined as in PXX.  It is clear that it is the parameter
$\beta$ that we constrain in our analysis. 

Now, PXX parameterizes the scaling relation between mass and
$Y_{\rm SZ}$ --- which we hereby denote $Y_b$ for reasons that will be made apparent momentarily ---
via 
\be
Y_{b} = A(z)\left( \frac{(1-b)\Mfive}{6\times 10^{14}\ \msun} \right)^{\alpha}.
\ee
The function $A(z)$ is a known function of redshift whose specific form is irrelevant for this work.
Likewise, $\alpha$ is some known constant whose value is irrelevant for this work.
For the discussion below, it will be convenient to define the mass $M_b \equiv (1-b)\Mfive$, so
that
\be
Y_b = A\left( \frac{M_b}{6\times 10^{14}\ \msun} \right)^{\alpha}. \label{eq:Yb}
\ee

Now, the detection probability of a halo of mass $\Mfive$ in PXX is computed as follows: given $b$,
one first computes $M_b$, which is to be thought of as an X-ray hydrostatic mass proxy.  Since the mass
dependence of the templates used by \Planck\ are calibrated on X-ray hydrostatic masses, these templates
are defined in terms of $M_b$.  That is, the integrated $Y$ profile can be expressed via
\be
Y(R) = Y_b f(R/R_b) \label{eq:profile}
\ee
where $f(1)=1$, and $f$ is a known function. 

We contrast the above templates to how the observable $\Msz$ is defined.  Given an integrated SZ profile
$Y(R)$, the mass $\Msz$ is given by the solution to the equation 
\be
Y(R_{SZ}) = A \left( \frac{\Msz}{6\times 10^{14}\ \msun} \right)^{\alpha}
\ee
where $R_{SZ}$ is the radius of a cluster of mass $M_{SZ}$.  

Let us then assume that a cluster of mass $M_{SZ}$ corresponds to a cluster of mass $M_b$,
so that the profile $Y(R)$ is given by equation~\ref{eq:profile}.  The mass $M_{SZ}$ is given
by the solution to:
\be
Y_b f(R_{SZ}/R_b) = A \left( \frac{\Msz}{6\times 10^{14}\ \msun} \right)^{\alpha}.
\ee
Setting $\Msz=M_b$, one has then that $f=1$, while the right hand side reduces to the right
hand side of equation~\ref{eq:Yb}.  The equality is valid, and therefore $\Msz=M_b$ is precisely
the solution we were looking for.  Since $\Msz=M_b$, it follows that $\beta=b$, and therefore
our analysis is directly relevant to the \Planck\ cosmological analysis without the need to introduce
aperture corrections.

\end{document}